\documentclass[aps,prc,preprint,superscriptaddress,nofootinbib,showpacs,showkeys]{revtex4-1}
\usepackage{amsmath}
\usepackage{amssymb}
\usepackage{bm}% bold math
\usepackage{braket}% braket
\usepackage{color}% Color \textcolor[rgb]{0,0.833,1}{}
\usepackage{dcolumn}% Align table columns on decimal point
\usepackage{graphicx}% Include figure files
\usepackage{hyperref}% add hypertext capabilities

\begin{document}
% \title{Fock contribution to nuclear symmetry energy and its slope parameter in relativistic theory}
\title{Fock contributions to nuclear symmetry energy and its slope parameter based on Lorentz-covariant decomposition of nucleon self-energies}
% \title{Fock contribution on nuclear symmetry energy and its slope parameter in a relativistic framework}
% \title{Exchange contribution in Fock terms on nuclear symmetry energy and its slope parameter in a relativistic framework}
% \title{Fock contribution on the properties of nuclear matter in a relativistic framework}
% \title{Fock contribution on nuclear symmetry energy in a relativistic framework}
% \title{Role of Fock terms on nuclear symmetry energy in a relativistic framework}
%%
\author{Tsuyoshi Miyatsu}
\email[]{tsuyoshi.miyatsu@rs.tus.ac.jp}
\affiliation{Department of Physics, Faculty of Science and Technology, Tokyo University of Science, Noda 278-8510, Japan}
\author{Myung-Ki Cheoun}
\email[]{cheoun@ssu.ac.kr}
% \affiliation{Department of Physics, Soongsil University, Seoul 156-743, Korea}
\affiliation{Department of Physics and Origin of Matter and Evolution of the Galaxies (OMEG) Institute, Soongsil University, Seoul 156-743, Korea}
\author{Chikako Ishizuka}
% \email[]{}
% \affiliation{Research Laboratory for Nuclear Reactors, Tokyo Institute of Technology, Tokyo, Japan}
\affiliation{Laboratory for Advanced Nuclear Energy, Institute of Innovative Research, Tokyo Institute of Technology, Tokyo, 152-8550 Japan}
%%
% \author{Kyungsik Kim}
\author{K.~S.~Kim}
% \email[]{}
\affiliation{School of Liberal Arts and Science, Korea Aerospace University, Goyang 412-791, Korea}
\author{Tomoyuki Maruyama}
% \email[]{}
\affiliation{College of Bioresource Sciences, Nihon University, Fujisawa 252-8510, Japan}
\author{Koichi Saito}
\email[]{koichi.saito@rs.tus.ac.jp}
\affiliation{Department of Physics, Faculty of Science and Technology, Tokyo University of Science, Noda 278-8510, Japan}
% \affiliation{Department of Physics, Faculty of Science and Technology, Tokyo University of Science, Noda 278-8510, Japan \ }
%%
\date{\today}
\begin{abstract}
Using relativistic Hartree-Fock (RHF) approximation, we study the effect of Fock terms on the nuclear properties not only around the saturation density, $\rho_{0}$, but also at higher densities.
In particular, we investigate how the momentum dependence due to the exchange contribution affects the nuclear symmetry energy and its slope parameter, using the Lorentz-covariant decomposition of nucleon self-energies in an extended version of the RHF model, in which the exchange terms are adjusted so as to reproduce the single-nucleon potential at $\rho_{0}$.
We find that the Fock contribution suppresses the kinetic term of nuclear symmetry energy at the densities around and beyond $\rho_{0}$.
It is noticeable that not only the isovector-vector ($\rho$) meson but also the isoscalar mesons ($\sigma, \omega$) and pion make significant influence on the potential term of nuclear symmetry energy through the exchange diagrams.
Furthermore, the exchange contribution prevents the slope parameter from increasing monotonically at high densities.
\end{abstract}
\maketitle

%%%%%%%%%%%%%%%%%%%%%%%%%%%%%%%%%%%%%%%%%%%%%%%%%%%%%%%%%%%%%%%%%%%%%%%%%%%%%%%%%%%%%%%%%%%%%%%%%%%%
\section{Introduction}
\label{sec:introduction}
%%%%%%%%%%%%%%%%%%%%%%%%%%%%%%%%%%%%%%%%%%%%%%%%%%%%%%%%%%%%%%%%%%%%%%%%%%%%%%%%%%%%%%%%%%%%%%%%%%%%

The nuclear symmetry energy, $E_{\rm sym}$, which is defined as the difference between the energies of pure neutron and symmetric nuclear matter, is recognized to be an important physical quantity in nuclear physics and astrophysics~\cite{Lattimer:2014scr,Li:2008gp}.
It can account for many experimental facts at low nuclear densities, especially the existence of neutron skin and the different distributions of neutrons and protons in a nucleus~\cite{Danielewicz:2008cm,Danielewicz:2013upa,Danielewicz:2016bgb,Chen:2007ih,Choi:2017afh}.
It also plays an important role to explain the properties of isospin-asymmetric nuclear matter in the density region beyond the saturation density, $\rho_0$, experimentally realized by the heavy-ion collisions~\cite{Russotto:2011hq,Russotto:2016ucm,Danielewicz:2002pu,Fuchs:2005zg,Li:2005jy,Klahn:2006ir,Tsang:2008fd}.
At the same time, some astrophysical observations, for instance the mass-radius relations of neutron stars and the cooling process of proto-neutron stars, strongly depend on $E_{\rm sym}$~\cite{Lattimer:2012xj,Lattimer:2012nd,Kolomeitsev:2016sjl,Oertel:2016bki}.

So far, many theoretical discussions on the properties of symmetric and asymmetric nuclear matter have been performed.
The phenomenological calculations based on the effective many-body interaction with the Skyrme or Gogny force have been adopted for a long time to examine the structure of finite nuclei and infinite nuclear matter~\cite{RikovskaStone:2003bi,Dutra:2012mb,Decharge:1979fa}.
Recently, the relativistic mean-field (RMF) models based on Quantum Hadrodynamics (QHD) have been widely applied to study astrophysical phenomena as well as the properties of nuclear matter at low densities~\cite{Glendenning:2000,Glendenning:2001pe,Shen:1998gq,Shen:2011qu,Ishizuka:2008gr}.
Meanwhile, microscopic studies, for example the so-called Dirac-Brueckner-Hartree-Fock (DBHF) approach~\cite{Brockmann:1990cn,Katayama:2013zya,Katayama:2015dga} and chiral perturbation theory~\cite{Scherer:2002tk,Entem:2003ft,Machleidt:2011zz} based on realistic $NN$ interactions, have also succeeded in the description of nuclear saturation features.
However, depending on the models, different results of the high-dense behavior of $E_{\rm sym}$ have been reported so far, and, in particular, its density dependence beyond $\rho_0$ is undetermined yet~\cite{Tsang:2012se}.

Since the observations of massive neutron-stars and the hyperon puzzle in astrophysics have provided valuable information on the equation of state (EoS) for neutron-star matter, it helps us study $E_{\rm sym}$ at high densities~\cite{Lonardoni:2014bwa,Bednarek:2011gd,Bombaci:2016xzl}.
In addition, the recent tidal deformation data induced by the gravitational wave from binary neutron star merger detected by LIGO scientific and Virgo collaborations~\cite{TheLIGOScientific:2017qsa,Abbott:2018wiz} may be useful to make a constraint on the EoS.
Several models have been proposed recently, in which the astrophysical constraints as well as the terrestrial experimental data in nuclear physics are satisfied simultaneously.
Within relativistic calculations, those models can be classified into two categories.
One is calculated in relativistic Hartree (RH) approximation with additional repulsive force using SU(3) flavor symmetry, or nonlinear terms with multi-meson couplings~\cite{Miyatsu:2013yta,Miyatsu:2014wca,Weissenborn:2011ut,Lopes:2013cpa,Zhao:2014rra,Oertel:2014qza,Gomes:2017zkc,Tsubakihara:2012ic,Maslov:2015msa,Maslov:2015wba}.
The other is based on relativistic Hartree-Fock (RHF) approximation~\cite{Miyatsu:2011bc,Katayama:2012ge,Miyatsu:2015kwa,Miyatsu:2013hea,Whittenbury:2013wma,Thomas:2013sea,Whittenbury:2015ziz,Li:2018jvz,Li:2018qaw}.
Since only the direct diagram is considered in the RH model, the exchanging mesons take zero momentum transfer.
In contrast, the inclusion of Fock terms automatically allows us to explore the momentum dependence at meson-nucleon vertices.
In addition, the pion contribution can be taken into account through the exchange diagram.
However, it has not yet been investigated in detail how the exchange terms affect the properties of dense matter.
Especially, we should discuss the effect of nucleon Fermi velocity on $E_{\rm sym}$ because the momentum dependence largely changes the Fermi velocity without varying the other properties of nuclear matter~\cite{Maruyama:1999ye,Maruyama:2005iw}.

Recent experimental analyses on the density dependence of $E_{\rm sym}$ have been performed by using the free Fermi Gas model, in which $E_{\rm sym}$ can be separated into the kinetic and potential terms as $E_{\rm sym}(\rho_{B})=E_{\rm sym}^{\rm kin}(\rho_{B})+E_{\rm sym}^{\rm pot}(\rho_{B})$,  where $\rho_{B}$ is the total baryon  density~\cite{Tsang:2008fd,Zhang:2007qv,Sammarruca:2017edz,Hen:2014yfa,Dutra:2017ysk}.
In this decomposition, $E_{\rm sym}^{\rm kin}$ is expressed in terms of only the degrees of freedom of nucleons, and it is typically parametrized as $E_{\rm sym}^{\rm kin}(\rho_{0})(\rho_{B}/\rho_{0})^{2/3}$, where $E_{\rm sym}^{\rm kin}(\rho_{0})$ is the symmetry-energy value at $\rho_{0}$. Meanwhile, $E_{\rm sym}^{\rm pot}$ is simply given by the power-low function as  $(\rho_{B}/\rho_{0})^{\gamma}$ with a parameter $\gamma$.

On the other hand, some theoretical calculations on $E_{\rm sym}$ have been carried out  in the RHF model~\cite{Sun:2008zzk,Zhao:2014bga,Sun:2016ley,Liu:2018far}, and $E_{\rm sym}$ is again often divided into $E_{\rm sym}^{\rm kin}$ and $E_{\rm sym}^{\rm pot}$, which are expressed in terms of the derivatives of the expectation valuables, $\braket{T}$ and $\braket{V}$, where the nuclear matter Hamiltonian is given as $H=T+V$~\cite{Vidana:2011ap}.
Since this decomposition does not correspond to that in the experimental analyses, it may be misleading to compare the components, $E_{\rm sym}^{\rm kin}$ and $E_{\rm sym}^{\rm pot}$, in the experimental results with the theoretical results calculated in Refs.~\cite{Sun:2008zzk,Zhao:2014bga,Sun:2016ley,Liu:2018far}.

In contrast, an alternative decomposition based on the Lorentz-covariant forms of nucleon self-energies has been proposed in Refs.~\cite{Czerski:2002pz,Cai:2012en,Liu:2018kfv}, and it is more consistent with the experimental analyses.
In the present paper, we adopt this decomposition to clarify the density dependence of $E_{\rm sym}$ and its slope parameter, $L$, and study the properties of dense matter in detail within RHF approximation~\cite{Serot:1984ey,Bouyssy:1987sh}.
The effect of Fock contributions, namely the momentum dependence of nucleon self-energies, is very important to understand the properties of asymmetric nuclear matter, including $E_{\rm sym}$ and $L$.

This paper is organized as follows.
In Section~\ref{sec:theoretical-formalism}, a brief review for the RHF formalism based on QHD is presented.
The analytical derivation of $E_{\rm sym}$ and $L$ is then demonstrated.
Numerical results and discussions are addressed in Section~\ref{sec:results}.
We here study the nuclear matter properties from point of view of the Fock terms.
In particular, we investigate the momentum dependence of $E_{\rm sym}$ and $L$ at high densities, and compare our results with the recent experimental data, focusing on the $\gamma$ parameter in $E_{\rm sym}^{\rm pot}$.
Finally, we give a summary in Section~\ref{sec:summary}.

%%%%%%%%%%%%%%%%%%%%%%%%%%%%%%%%%%%%%%%%%%%%%%%%%%%%%%%%%%%%%%%%%%%%%%%%%%%%%%%%%%%%%%%%%%%%%%%%%%%%
\section{Theoretical formalism}
\label{sec:theoretical-formalism}
%%%%%%%%%%%%%%%%%%%%%%%%%%%%%%%%%%%%%%%%%%%%%%%%%%%%%%%%%%%%%%%%%%%%%%%%%%%%%%%%%%%%%%%%%%%%%%%%%%%%

%%%%%%%%%%%%%%%%%%%%%%%%%%%%%%%%%%%%%%%%%%%%%%%%%%%%%%%%%%%%%%%%%%%%%%%%%%%%%%%%
\subsection{Relativistic Hartree-Fock approach}
\label{subsec:RHF}
%%%%%%%%%%%%%%%%%%%%%%%%%%%%%%%%%%%%%%%%%%%%%%%%%%%%%%%%%%%%%%%%%%%%%%%%%%%%%%%%

For the description of uniform nuclear matter, we present the relativistic formulation based on the QHD model in Hartree-Fock approximation.
The total Lagrangian density is written as~\cite{Serot:1984ey,Bouyssy:1987sh}
\begin{align}
  \mathcal{L}
  &= \sum_{N=p,n}\bar{\psi}_{N}\left(i\gamma_{\mu}\partial^{\mu}-M_{N}\right)\psi_{N} + \mathcal{L}_{M} + \mathcal{L}_{\rm int} -U_{\rm NL},
\label{eq:Lagrangian-total}
\end{align}
where $\psi_{N}$ is the nucleon $(N)$ field with the mass in vacuum, $M_{N}=939$ MeV.

The meson term reads
\begin{widetext}
\begin{align}
  \mathcal{L}_{M}
  &= \frac{1}{2}\left(\partial_{\mu}\sigma\partial^{\mu}\sigma-m_{\sigma}^{2}\sigma^{2}\right)
  +  \frac{1}{2}m_{\omega}^{2}\omega_{\mu}\omega^{\mu} - \frac{1}{4}W_{\mu\nu}W^{\mu\nu}
  \nonumber \\
  &+ \frac{1}{2}m_{\rho}^{2}\bm{\rho}_{\mu}\cdot\bm{\rho}^{\mu} - \frac{1}{4}\bm{R}_{\mu\nu}\cdot\bm{R}^{\mu\nu}
  +  \frac{1}{2}\left(\partial_{\mu}\bm{\pi}\cdot\partial^{\mu}\bm{\pi}-m_{\pi}^{2}\bm{\pi}^{2}\right),
  \label{eq:Lagrangian-meson}
\end{align}
\end{widetext}
with $W_{\mu\nu}$ ($\bm{R}_{\mu\nu}$) being the field strength tensor for $\omega$ ($\bm{\rho}$) meson.
The meson masses are respectively chosen as $m_{\sigma}=550$ MeV, $m_{\omega}=783$ MeV, $m_{\rho}=770$ MeV, and $m_{\pi}=138$ MeV.

The interaction Lagrangian density is given by
\begin{widetext}
\begin{align}
  \mathcal{L}_{\rm int}
  &= \sum_{N=p,n}\bar{\psi}_{N} \biggl( g_{\sigma}\sigma
  -  g_{\omega}\gamma_{\mu}\omega^{\mu}
  -  g_{\rho}\gamma_{\mu}\bm{\rho}^{\mu}\cdot\bm{\tau}_{N}
  \nonumber \\
  &\hspace{2.0cm}
  + \frac{f_{\rho}}{2\mathcal{M}}\sigma_{\mu\nu}\partial^{\nu}\bm{\rho}^{\mu}\cdot\bm{\tau}_{N}
  -  \frac{f_{\pi}}{m_{\pi}}\gamma_{5}\gamma_{\mu}\partial^{\mu}\bm{\pi}\cdot\bm{\tau}_{N} \biggr) \psi_{N},
  \label{eq:Lagrangian-interaction}
\end{align}
\end{widetext}
where $\bm{\tau}_{N}$ is the isospin matrix for nucleon, and $\sigma_{\mu\nu}=\frac{i}{2}\left[\gamma_{\mu},\gamma_{\nu}\right]$.
The $\sigma$-, $\omega$-, $\rho$-, and $\pi$-$N$ coupling constants are respectively denoted by $g_{\sigma}$, $g_{\omega}$, $g_{\rho}$ and $f_{\pi}$, while $f_{\rho}$ is the tensor coupling constant for $\bm{\rho}$ meson.
In the present calculation, the tensor coupling for $\omega$ meson is neglected, since the $\omega$-$N$ tensor coupling constant is small~\cite{Bouyssy:1987sh,Brockmann:1990cn}.

In order to obtain a quantitative description of the nuclear ground-state properties, the following nonlinear potential for $\sigma$ meson is, at least, required to consider~\cite{Boguta:1977xi}:
\begin{equation}
  U_{\rm NL} = \frac{1}{3}g_{2}\sigma^{3} + \frac{1}{4}g_{3}\sigma^{4}.
  \label{eq:Lagrangian-NL}
\end{equation}
Since, in RHF approximation, the precise treatment of nonlinear terms involves tremendous difficulties~\cite{Massot:2008pf,Massot:2009kk}, we simply replace the $\sigma$ field in Eq.~\eqref{eq:Lagrangian-NL} by its ground-state expectation value, $\bar{\sigma}$, in the present calculation~\cite{Bernardos:1993re,Weber:1999qn}.

To sum up all orders of the tadpole (Hartree) and exchange (Fock) diagrams in the nucleon Green's function, $G_{N}$, we consider the Dyson's equation
\begin{equation}
  G_{N}(k) = G_{N}^{0}(k) + G_{N}^{0}(k)\Sigma_{N}(k)G_{N}(k),
  \label{eq:Dyson-equation}
\end{equation}
where $k^\mu$ is the four momentum of nucleon, $\Sigma_{N}$ is the nucleon self-energy, and $G_{N}^{0}$ is the nucleon Green's function in free space.
The nucleon self-energy in matter can be written as \citep{Serot:1984ey}
\begin{equation}
  \Sigma_{N}(k) = \Sigma_{N}^{s}(k) - \gamma_{0}\Sigma_{N}^{0}(k) + (\bm{\gamma}\cdot\hat{k})\Sigma_{N}^{v}(k),
  \label{eq:nucleon-self-engy}
\end{equation}
with $\hat{k}$ being the unit vector along the (three) nucleon momentum, $\bm{k}$.
It can be divided into the scalar ($s$), time ($0$), and space ($v$) components, which provide the effective nucleon mass, momentum, and energy in matter~\cite{Miyatsu:2011bc,Katayama:2012ge,Miyatsu:2015kwa}:
\begin{align}
  & M_{N}^{\ast}(k) = M_{N} + \Sigma_{N}^{s}(k),
  \label{eq:auxiliary-quantity-mass} \\
  & k_{N}^{\ast\mu} = (k_{N}^{\ast0},\bm{k}_{N}^{\ast}) = (k^{0}+\Sigma_{N}^{0}(k),\bm{k}+\hat{k}\Sigma_{N}^{v}(k)),
  \label{eq:auxiliary-quantity-momentum} \\
  & E_{N}^{\ast}(k) = \left[\bm{k}_{N}^{\ast2}+M_{N}^{\ast2}(k)\right]^{1/2}.
  \label{eq:auxiliary-quantity-energy}
\end{align}
In the mean-field approximation, the meson fields are replaced by their constant expectation values: $\bar{\sigma}$, $\bar{\omega}$, and $\bar{\rho}$ (the $\rho^{0}$ field).
The pion field vanishes in RH approximation, but it can be taken into account as the exchange contribution in RHF approximation.
In the present calculation, the retardation effect in Fock terms is ignored, since it gives at most a few percent contribution to the nucleon self-energies \citep{Serot:1984ey,Katayama:2012ge,Whittenbury:2013wma}.
All the components of nucleon self-energies, $\Sigma_{N}^{s,0,v}$, in Eq.~\eqref{eq:nucleon-self-engy} are then calculated by~\cite{Miyatsu:2011bc,Katayama:2012ge,Miyatsu:2015kwa}
\begin{widetext}
\begin{align}
  \Sigma^{s}_{N}(k)
  &= -g_{\sigma}\bar{\sigma}
  +  \sum_{N^{\prime}=p,n}\sum_{i=\sigma,\omega,\rho,\pi}\frac{\left(\tau_{iNN^{\prime}}\right)^{2}}{\left(4\pi\right)^{2}k}\int_{0}^{k_{F_{N^{\prime}}}}dq\,q
     \left[\frac{M^{\ast}_{N^{\prime}}(q)}{E^{\ast}_{N^{\prime}}(q)}B_{i}(k,q)
  +  \frac{q^{\ast}_{N^{\prime}}}{E^{\ast}_{N^{\prime}}(q)}D_{i}(q,k)\right],
  \label{eq:self-engy-scalar} \\
  \Sigma^{0}_{N}(k)
  &= -g_{\omega}\bar{\omega} - g_{\rho} (\bm{\tau}_{N})_{3}\bar{\rho}
     -\sum_{N^{\prime}=p,n}\sum_{i=\sigma,\omega,\rho,\pi}\frac{\left(\tau_{iNN^{\prime}}\right)^{2}}{\left(4\pi\right)^{2}k}\int_{0}^{k_{F_{N^{\prime}}}}dq\,qA_{i}(k,q),
  \label{eq:self-engy-time} \\
  \Sigma^{v}_{N}(k)
  &= \sum_{N^{\prime}=p,n}\sum_{i=\sigma,\omega,\rho,\pi}\frac{\left(\tau_{iNN^{\prime}}\right)^{2}}{\left(4\pi\right)^{2}k}\int_{0}^{k_{F_{N^{\prime}}}}dq\,q
     \left[\frac{q^{\ast}_{N^{\prime}}}{E^{\ast}_{N^{\prime}}(q)}C_{i}(k,q)
  +  \frac{M^{\ast}_{N^{\prime}}(q)}{E^{\ast}_{N^{\prime}}(q)}D_{i}(k,q)\right],
  \label{eq:self-engy-space}
\end{align}
\end{widetext}
with $k=|\bm{k}|$ and $q=|\bm{q}|$.
The $k_{F_{N}}$ is the Fermi momentum for nucleon $N$, and the factor, $\tau_{iNN^{\prime}}$, is the isospin weight at meson-$NN^{\prime}$ vertex in the exchange diagrams.
In addition, the functions $A_{i}$, $B_{i}$, $C_{i}$, and $D_{i}$ are explicitly specified in Table~\ref{tab:NSE}, in which the following functions are used \citep{Katayama:2012ge,Miyatsu:2015kwa}:
\begin{widetext}
\begin{align}
  \Theta_{i}(k,q) &= \frac{\Lambda_{i}^{8}}{(m_{i}^{2}-\Lambda_{i}^{2})^{4}}
                     \left(\ln\left[\frac{M_i^{+}(k,q)}{M_i^{-}(k,q)}\frac{L_i^{-}(k,q)}{L_i^{+}(k,q)}\right]
                   + \sum_{n=1}^{3}\left(m_{i}^{2}-\Lambda_{i}^{2}\right)^{n}N_{i}^{n}(k,q)\right),
  \label{eq:Theta-func} \\
  \Phi_{i}(k,q)   &= \frac{1}{4kq}\left[\left(k^{2}+q^{2}+m_{i}^{2}\right)\Theta_{i}(k,q) - \Lambda_{i}^{8}N_{i}^{3}(k,q)\right],
  \label{eq:Phi-func} \\
  \Psi_{i}(k,q)   &= \left(k^{2}+q^{2}-m_{i}^{2}/2\right)\Phi_{i}(k,q) -kq\Theta_{i}(k,q) + \Omega_{i}(k,q),
  \label{eq:Psi-func} \\
  \Pi_{i}(k,q)    &= \left(k^{2}+q^{2}\right)\Phi_{i}(k,q) - kq\Theta_{i}(k,q) + \Omega_{i}(k,q),
  \label{eq:Pi-func} \\
  \Gamma_{i}(k,q) &= \left[k\Theta_{i}(k,q)-2q\Phi_{i}(k,q)\right],
  \label{eq:Gamma-func}
\end{align}
\end{widetext}
where
\begin{widetext}
\begin{align}
  \Omega_{i}(k,q)  &= \frac{\Lambda_{i}^{8}}{4kq}\left[N_{i}^{2}(k,q)+\left(k^{2}+q^{2}+\Lambda_{i}^{2}\right)N_{i}^{3}(k,q)\right],
  \label{eq:Omega-func}  \\
  L_{i}^{\pm}(k,q) &= \Lambda_{i}^{2} + (k \pm q)^{2},
  \label{eq:L-func} \\
  M_{i}^{\pm}(k,q) &= m_{i}^{2} + (k \pm q)^{2},
  \label{eq:M-func} \\
  N_{i}^{n}(k,q)   &= \frac{(-1)^{n}}{n}\left(\left[L_{i}^{+}(k,q)\right]^{-n}-\left[L_{i}^{-}(k,q)\right]^{-n}\right),
  \label{eq:N-func}
\end{align}
\end{widetext}
%
%%%%%%%%%%%%%%%%%%%%%%%%%%%%%%%%%%%%%%%%%%%%%%%%%%%%%%%%%%%%%%%%%%%%%%%%%%%%%%%%
%% Table: NSE
%%%%%%%%%%%%%%%%%%%%%%%%%%%%%%%%%%%%%%%%%%%%%%%%%%%%%%%%%%%%%%%%%%%%%%%%%%%%%%%%
\begin{table*}
\caption{\label{tab:NSE}
Functions $A_{i}$, $B_{i}$, $C_{i}$, and $D_{i}$.
The index $i$ is specified in the left column, where $V (T)$ stands for the vector (tensor) coupling at meson-$NN^{\prime}$ vertex.
The bottom row is for the (pseudovector) pion contribution.
The functions, $\Theta_{i}$, $\Phi_{i}$, $\Psi_{i}$, $\Pi_{i}$, and $\Gamma_{i}$, are given in the text.
}
\begin{ruledtabular}
\begin{tabular}{lllll}
$i$ & $A_{i}$ & $B_{i}$ & $C_{i}$ & $D_{i}$ \\
\hline
$\sigma$ &
$g_{\sigma}^{2}\Theta_{\sigma}$ &
$g_{\sigma}^{2}\Theta_{\sigma}$ &
$-2g_{\sigma}^{2}\Phi_{\sigma}$ &
-- \\
$\omega$ &
$2g_{\omega}^{2}\Theta_{\omega}$ &
$-4g_{\omega}^{2}\Theta_{\omega}$ &
$-4g_{\omega}^{2}\Phi_{\omega}$ &
-- \\
$\rho_{VV}$ &
$2g_{\rho}^{2}\Theta_{\rho}$ &
$-4g_{\rho}^{2}\Theta_{\rho}$ &
$-4g_{\rho}^{2}\Phi_{\rho}$ &
-- \\
$\rho_{TT}$ &
$-\left(\frac{f_{\rho}}{2\mathcal{M}}\right)^{2}m_{\rho}^{2}\Theta_{\rho}$ &
$-3\left(\frac{f_{\rho}}{2\mathcal{M}}\right)^{2}m_{\rho}^{2}\Theta_{\rho}$ &
$4\left(\frac{f_{\rho}}{2\mathcal{M}}\right)^{2}\Psi_{\rho}$ &
-- \\
$\rho_{VT}$ &
-- &
-- &
-- &
$6\frac{f_{\rho}g_{\rho}}{2\mathcal{M}}\Gamma_{\rho}$ \\
$\pi_{pv}$ &
$-f_{\pi}^{2}\Theta_{\pi}$ &
$-f_{\pi}^{2}\Theta_{\pi}$ &
$2\left(\frac{f_{\pi}}{m_{\pi}}\right)^{2}\Pi_{\pi}$ &
-- \\
\end{tabular}
\end{ruledtabular}
\end{table*}
%%%%%%%%%%%%%%%%%%%%%%%%%%%%%%%%%%%%%%%%%%%%%%%%%%%%%%%%%%%%%%%%%%%%%%%%%%%%%%%%
with $\Lambda_{i}$ being a cutoff parameter at interaction vertex specified by $i$, as shown in the 1st column of Table~\ref{tab:NSE}.
In the present calculation, a dipole-type form factor is introduced at each interaction vertex~\cite{Miyatsu:2011bc,Katayama:2012ge,Miyatsu:2015kwa}:
\begin{equation}
  F_{i}(\left(\bm{k}-\bm{q}\right)^{2}) = \left[\frac{\Lambda_{i}^{2}}{\Lambda_{i}^{2}+\left(\bm{k}-\bm{q}\right)^{2}}\right]^{2}.
  \label{eq:form-factor}
\end{equation}
We here employ $\Lambda_{\sigma}=2.0$ GeV, $\Lambda_{\omega}=1.5$ GeV, $\Lambda_{\rho}=1.3$ GeV, and $\Lambda_{\pi}=1.2$ GeV~\cite{Brockmann:1990cn}, and the effect of the form factor vanishes in the limit, $\Lambda_{i}\to\infty$.

As in the case of the RH model, the mean-field values of $\bar{\sigma}$, $\bar{\omega}$, and $\bar{\rho}$ in Eqs.~\eqref{eq:self-engy-scalar} and  \eqref{eq:self-engy-time} are given by
\begin{align}
  \bar{\sigma} &= \sum_{N=p,n}\frac{g_{\sigma}}{m_{\sigma}^{2}}\rho_{N}^{s}
                - \frac{1}{m_{\sigma}^{2}}\left(g_{2}\bar{\sigma}^{2}+g_{3}\bar{\sigma}^{3}\right),
  \label{eq:EOM-sigma} \\
  \bar{\omega} &= \sum_{N=p,n}\frac{g_{\omega}}{m_{\omega}^{2}}\rho_{N},
  \label{eq:EOM-omega} \\
  \bar{\rho}   &= \sum_{N=p,n}\frac{g_{\rho}}{m_{\rho}^{2}}(\bm{\tau}_{N})_{3}\rho_{N},
  \label{eq:EOM-rho}
\end{align}
where the scalar and nucleon densities are, respectively, written as
\begin{align}
  \rho_{N}^{s} &= \frac{1}{\pi^{2}}\int_{0}^{k_{F_{N}}}dk~k^{2}\frac{M_{N}^{\ast}(k)}{E_{N}^{\ast}(k)},
  \label{eq:nucleon-scalar-density} \\
  \rho_{N}     &= \frac{1}{\pi^{2}}\int_{0}^{k_{F_{N}}}dk~k^{2} = \frac{k_{F_{N}}^{3}}{3\pi^{2}}.
\end{align}

Once the nucleon self-energies shown in Eqs.~\eqref{eq:self-engy-scalar}--\eqref{eq:self-engy-space} are calculated, the total energy density for uniform nuclear matter is determined by the energy-momentum tensor.
It can be given by a sum of the kinetic and potential terms of nucleon and the nonlinear term,
\begin{equation}
  \epsilon = \epsilon_{\rm nucl}^{\rm kin} + \epsilon_{\rm nucl}^{\rm pot} + \epsilon_{\rm NL},
  \label{eq:engy-density}
\end{equation}
with
\begin{widetext}
\begin{align}
  \epsilon_{\rm nucl}^{\rm kin} &= \sum_{N=p,n} \frac{1}{\pi^{2}} \int_{0}^{k_{F_{N}}} dk \, k^{2} E_{N}^{\ast}(k),
  \label{eq:engy-density-kin} \\
  \epsilon_{\rm nucl}^{\rm pot} &= -\sum_{N=p,n} \frac{1}{2\pi^{2}} \int_{0}^{k_{F_{N}}} dk \, k^{2}
                                    \left[\frac{\Sigma_{N}^{s}(k)M_{N}^{\ast}(k)}{E_{N}^{\ast}(k)} + \Sigma_{N}^{0}(k)
                                 +  \frac{\Sigma_{N}^{v}(k)k_{N}^{\ast}(k)}{E_{N}^{\ast}(k)}\right],
  \label{eq:engy-density-pot} \\
  \epsilon_{\rm NL}             &= -\frac{1}{2}\left(\frac{1}{3}g_{2}\bar{\sigma}^{3}+\frac{1}{2}g_{3}\bar{\sigma}^{4}\right).
  \label{eq:engy-density-NL}
\end{align}
\end{widetext}
The pressure for uniform matter is then obtained from thermodynamics relation,
\begin{equation}
  P = \rho_{B}^{2}\frac{\partial}{\partial\rho_{B}}\left(\frac{\epsilon}{\rho_{B}}\right).
  \label{eq:pressure}
\end{equation}
%

%%%%%%%%%%%%%%%%%%%%%%%%%%%%%%%%%%%%%%%%%%%%%%%%%%%%%%%%%%%%%%%%%%%%%%%%%%%%%%%%
\subsection{Symmetry energy and its slope parameter}
\label{subsec:Esym-L-def}
%%%%%%%%%%%%%%%%%%%%%%%%%%%%%%%%%%%%%%%%%%%%%%%%%%%%%%%%%%%%%%%%%%%%%%%%%%%%%%%%

Using the Hugenholtz--Van Hove theorem, the nuclear symmetry energy is generally derived by~\cite{Czerski:2002pz,Cai:2012en}
\begin{equation}
  E_{\rm sym} = E_{\rm sym}^{\rm kin} + E_{\rm sym}^{\rm pot},
  \label{eq:Esym}
\end{equation}
with $E_{\rm sym}^{\rm kin (pot)}$ being the kinetic (potential) term for $E_{\rm sym}$.
Using the nucleon self-energies, the $E_{\rm sym}^{\rm kin}$ and $E_{\rm sym}^{\rm pot}$ in RHF approximation are respectively given by
\begin{align}
  E_{\mathrm{sym}}^{\rm kin} &= \frac{1}{6}\frac{k_{F}^{\ast}}{E_{F}^{\ast}}k_{F},
  \label{eq:Esym-kin} \\
  E_{\mathrm{sym}}^{\rm pot} &= \frac{1}{8}\rho_{B}\left(\frac{M_{N}^{\ast}}{E_{F}^{\ast}}\partial\Sigma_{\rm sym}^{s}
                              - \partial\Sigma_{\rm sym}^{0} + \frac{k_{F}^{\ast}}{E_{F}^{\ast}}\partial\Sigma_{\rm sym}^{v}\right),
  \label{eq:Esym-pot}
\end{align}
with $k_{F}=k_{F_{p}}=k_{F_{n}}$, $E_{F}^{\ast} = \sqrt{k_{F}^{\ast2}+M_{N}^{\ast2}}$, and
\begin{equation}
  \partial\Sigma_{\rm sym}^{s(0)[v]} \equiv \left(\frac{\partial}{\partial\rho_{p}}-\frac{\partial}{\partial\rho_{n}}\right)
                                            \left(\Sigma_{p}^{s(0)[v]}-\Sigma_{n}^{s(0)[v]}\right).
  \label{eq:Esym-func}
\end{equation}
We here note that the definitions of the kinetic and potential terms shown in Eqs.~\eqref{eq:Esym-kin} and \eqref{eq:Esym-pot} are different from those given in Refs.~\cite{Zhao:2014bga,Sun:2016ley,Liu:2018far}.
Since the nucleon self-energies, $\Sigma_{N}^{s,0,v}$, can be separated into the direct and exchange contributions, the direct one in $E_{\rm sym}^{\rm pot}$ is exactly the same as in RH approximation~\cite{Chen:2007ih,Dutra:2014qga}:
\begin{equation}
  E_{\mathrm{sym}}^{\rm pot, dir} = \frac{1}{2}\frac{g_{\rho}^{2}}{m_{\rho}^{2}}\rho_{B}.
  \label{eq:Esym-pot-dir}
\end{equation}

The slope parameter of nuclear symmetry energy, $L$, is also given by the kinetic and potential terms,
\begin{equation}
  L = L^{\rm kin} + L^{\rm pot},
  \label{eq:slope}
\end{equation}
with
\begin{widetext}
\begin{align}
  L^{\rm kin} &= \frac{1}{6}k_{F}\left[\frac{k_{F}^{\ast}}{E_{F}^{\ast}}
               + \frac{k_{F}}{E_{F}^{\ast}}\left(\frac{M_{N}^{\ast}}{E_{F}^{\ast}}\right)^{2}
               + \frac{2k_{F}^{2}}{\pi^{2}}\frac{k_{F}}{E_{F}^{\ast}}\frac{M_{N}^{\ast}}{E_{F}^{\ast}}\left(
                 \frac{M_{N}^{\ast}}{E_{F}^{\ast}}\frac{\partial\Sigma_{N}^{v}}{\partial\rho_{B}}
               - \frac{k_{F}^{\ast}}{E_{F}^{\ast}}\frac{\partial\Sigma_{N}^{s}}{\partial\rho_{B}}\right)\right],
  \label{eq:slope-kin} \\
  L^{\rm pot} &= 3E_{\rm sym}^{\rm pot}
               + \frac{3}{8}\rho_{B}\frac{\partial}{\partial\rho_{B}}\left(\frac{M_{N}^{\ast}}{E_{F}^{\ast}}\partial\Sigma_{\rm sym}^{s}
               - \partial\Sigma_{\rm sym}^{0} + \frac{k_{F}^{\ast}}{E_{F}^{\ast}}\partial\Sigma_{\rm sym}^{v}\right).
  \label{eq:slope-pot}
\end{align}
\end{widetext}
If we ignore the exchange contribution, that is, we take $k_{F}^{\ast}=k_{F}$, $\partial\Sigma_{N}^{s}/\partial\rho_{B}=\partial M_{N}^{\ast}/\partial\rho_{B}$, $\partial\Sigma_{N}^{v}/\partial\rho_{B}=0$, and $\partial\left(\partial\Sigma_{\rm sym}^{s(0)[v]}\right)/\partial\rho_{B}=0$, $L^{\rm kin}$ and $L^{\rm pot}$ are then equivalent to those in RH approximation. Thus, they are, respectively, written as~\cite{Dutra:2014qga}
\begin{align}
  L^{\rm kin, dir} &= \frac{1}{3}\frac{k_{F}^{2}}{\sqrt{k_{F}^{2}+M_{N}^{\ast2}}}
                      \left[1-\frac{k_{F}^{2}}{2\left(k_{F}^{2}+M_{N}^{\ast2}\right)}
                      \left(1+\frac{2M_{N}^{\ast}k_{F}}{\pi^{2}}\frac{\partial M_{N}^{\ast}}{\partial\rho_{B}}\right)\right],
  \label{eq:slope-kin-dir} \\
  L^{\rm pot, dir} &= \frac{3}{2}\frac{g_{\rho}^{2}}{m_{\rho}^{2}}\rho_{B}.
  \label{eq:slope-pot-dir}
\end{align}
%

%%%%%%%%%%%%%%%%%%%%%%%%%%%%%%%%%%%%%%%%%%%%%%%%%%%%%%%%%%%%%%%%%%%%%%%%%%%%%%%%%%%%%%%%%%%%%%%%%%%%
\section{Numerical results and discussions}
\label{sec:results}
%%%%%%%%%%%%%%%%%%%%%%%%%%%%%%%%%%%%%%%%%%%%%%%%%%%%%%%%%%%%%%%%%%%%%%%%%%%%%%%%%%%%%%%%%%%%%%%%%%%%

%%%%%%%%%%%%%%%%%%%%%%%%%%%%%%%%%%%%%%%%%%%%%%%%%%%%%%%%%%%%%%%%%%%%%%%%%%%%%%%%
\subsection{Determination of coupling constants}
\label{subsec:CCs}
%%%%%%%%%%%%%%%%%%%%%%%%%%%%%%%%%%%%%%%%%%%%%%%%%%%%%%%%%%%%%%%%%%%%%%%%%%%%%%%%

In the RMF model, the coupling constants are phenomenologically determined so as to reproduce the properties of infinite nuclear matter and finite nuclei~\cite{Lalazissis:2009zz,Sugahara:1993wz}.
In the present calculation, for simplicity, the coupling constants, $g_{\sigma}$, $g_{\omega}$, $g_{2}$, and $g_{3}$ in Eqs.~\eqref{eq:Lagrangian-interaction} and \eqref{eq:Lagrangian-NL}, are adjusted so as to fit the properties of symmetric nuclear matter at the saturation density, $\rho_{0}=0.16$ fm$^{-3}$, namely  the saturation energy ($-16.0$ MeV), the incompressibility $K_{0}$ ($=250$ MeV), and the effective nucleon mass ($M_{N}^{\ast}/M_{N}=0.70$).
These coupling constants are given in Table~\ref{tab:coupling-constants}, in which all the cases are calculated using RH or RHF approximation.

%%%%%%%%%%%%%%%%%%%%%%%%%%%%%%%%%%%%%%%%%%%%%%%%%%%%%%%%%%%%%%%%%%%%%%%%%%%%%%%%
%% Table: Coupling-constants
%%%%%%%%%%%%%%%%%%%%%%%%%%%%%%%%%%%%%%%%%%%%%%%%%%%%%%%%%%%%%%%%%%%%%%%%%%%%%%%%
\begin{table*}%[h!]
\caption{\label{tab:coupling-constants}
Coupling constants in various RMF calculations.
The coupling constant, $g_{\rho}$, in the model without asterisk is adjusted so as to fit the observed data, $E_{\rm sym}(\rho_{0})=32.5$ MeV, while that in the model with asterisk is taken to be $g_{\rho}^{2}/4\pi=0.55$.  In the bottom two rows, the coupling constants, $g_{M}$, appearing in the exchange terms are modified by the ratio, $w_{M}$, namely $g_{M} \to \tilde{g}_{M}=w_M g_{M}$, where $M=\sigma,\omega,\rho$ (see the text, for detail).
The coupling constants, $f_{\rho}$, and $f_{\pi}$, take the empirical values, $f_{\rho}/\tilde{g}_{\rho}=6.0$, and $f_{\pi}^{2}/4\pi=0.08$.
}
\begin{ruledtabular}
\begin{tabular}{lcccccccc}
Model        &           $g_{\sigma}$ &           $g_{\omega}$ &             $g_{\rho}$
             &           $w_{\sigma}$ &           $w_{\omega}$ &             $w_{\rho}$
             &    $g_{2}$ (fm$^{-1}$) &                $g_{3}$ \\
\hline
RH$^{\ast}$  &                   9.52 &                  10.36 &                   2.63
             & \multicolumn{1}{c}{--} & \multicolumn{1}{c}{--} & \multicolumn{1}{c}{--}
             &                  18.11 &               $-34.69$ \\
RH           &                   9.52 &                  10.36 &                   3.95
             & \multicolumn{1}{c}{--} & \multicolumn{1}{c}{--} & \multicolumn{1}{c}{--}
             &                  18.11 &               $-34.69$ \\
RHF$^{\ast}$ &                   6.57 &                   9.55 &                   2.63
             &                   1.00 &                   1.00 &                   1.00
             &                   8.92 &               $-38.17$ \\
RHF          &                   8.82 &                   9.08 &                   0.82
             &                   1.00 &                   1.00 &                   1.00
             &                  18.26 &               $-24.54$ \\
ERHF(low)    &                   8.07 &                   8.79 &                   2.63
             &                   1.00 &                   1.27 &                   0.40
             &                  19.90 &               $-31.48$ \\
ERHF(high)   &                   5.56 &                   6.06 &                   2.63
             &                   2.00 &                   2.88 &                   0.60
             &                  39.09 &               $ 83.74$ \\
\end{tabular}
\end{ruledtabular}
\end{table*}
%%%%%%%%%%%%%%%%%%%%%%%%%%%%%%%%%%%%%%%%%%%%%%%%%%%%%%%%%%%%%%%%%%%%%%%%%%%%%%%%
As for the $\rho$-$N$ coupling constants, $g_{\rho}$ and $f_{\rho}$, which are directly related to $E_{\rm sym}$, we consider the following two ways: (1) as shown in the RH$^{\ast}$ and RHF$^{\ast}$ models of Table~\ref{tab:coupling-constants}, we adopt the empirical values, $g_{\rho}^{2}/4\pi=0.55$ and $f_{\rho}/g_{\rho}\cong6$, which are suggested through the vector-meson-dominance model based on current algebra~\cite{Sakurai:1969,Hohler:1974ht,Hohler:1976ax}, (2) the coupling $g_{\rho}$ is chosen so as to satisfy the currently estimated value of $E_{\rm sym}$ at $\rho_{0}$, namely $E_{\rm sym}=32.5$ MeV, and the relation  $f_{\rho}/g_{\rho}\cong6$ is used in RHF approximation.
The models with this choice are denoted by RH and RHF (without asterisk) in Table~\ref{tab:coupling-constants}.

In addition, the pseudovector $\pi$-$N$ coupling constant is fixed as $f_{\pi}^{2}/4\pi=0.08$, derived from the low-energy $\pi N$ scattering data~\cite{deSwart:1990}.

Furthermore, we here consider an extended version of the RHF model (denoted by ERHF in Table~\ref{tab:coupling-constants}).
In the ERHF model, we replace the coupling constants, $g_{\sigma}$, $g_{\omega}$ and $g_{\rho}$, in the exchange terms with new ones, $\tilde{g}_{\sigma}$, $\tilde{g}_{\omega}$ and $\tilde{g}_{\rho}$, and the couplings in the direct terms remain unchanged (see Table~\ref{tab:coupling-constants}).
The purpose of this extension is to examine how the momentum dependence of nucleon self-energies contributes to various physical quantities through varying the strength of Fock terms~\cite{Weber:1992qc,Weber:1993et,Maruyama:1993jb}.
In this version, as in the RHF model, the coupling constants, $g_{\sigma}$, $g_{\omega}$, $g_{2}$, and $g_{3}$, are determined so as to fit the nuclear saturation properties, adopting the empirical values, $g_{\rho}^{2}/4\pi=0.55$, $f_{\rho}/\tilde{g}_{\rho}=6.0$, and $f_{\pi}^{2}/4\pi=0.08$.
The new coupling constants in the exchange terms, $\tilde{g}_{\sigma}$, $\tilde{g}_{\omega}$, and $\tilde{g}_{\rho}$, are then determined by simulating the empirical value of $E_{\rm sym}$ at $\rho_{0}$.
In Table~\ref{tab:coupling-constants}, we provide two parameter sets for the ERHF model, namely ERHF low (high), which can well reproduce the experimental data of single-nucleon potential at relatively low (high) kinetic energies.
This issue will be discussed in detail later.

%%%%%%%%%%%%%%%%%%%%%%%%%%%%%%%%%%%%%%%%%%%%%%%%%%%%%%%%%%%%%%%%%%%%%%%%%%%%%%%%
\subsection{Nucleon self-energy and Schr\"{o}dinger-equivalent potential}
\label{subsec:self-engy-USEP}
%%%%%%%%%%%%%%%%%%%%%%%%%%%%%%%%%%%%%%%%%%%%%%%%%%%%%%%%%%%%%%%%%%%%%%%%%%%%%%%%

In order to clarify the effect of Fock terms on the nuclear-matter properties, it is of importance to study the momentum dependence of nucleon self-energies, and the ERHF model may be one way to see it.

The Dirac optical model~\cite{Li:2013ck,Hama:1990vr} is useful to investigate the momentum dependence.
Although there are a lot of nonrelativistic single-nucleon potentials, in the present calculation, we consider the so-called Schr\"{o}dinger-equivalent potential (SEP) based on the Dirac equation with Lorentz-covariant scalar and vector self-energies for nucleon~\cite{Jaminon:1981xg}:
\begin{equation}
  U_{N}^{\rm SEP}(k,\epsilon_{k})
  = \Sigma_{N}^{s}(k) - \frac{E_{N}(k)}{M_{N}}\Sigma_{N}^{0}(k)
  + \frac{1}{2M_{N}}\left(\left[\Sigma_{N}^{s}(k)\right]^{2}
  - \left[\Sigma_{N}^{0}(k)\right]^{2}\right),
    \label{eq:USEP}
\end{equation}
where the nucleon kinetic energy, $\epsilon_{k}$, reads $\epsilon_{k}=E_{N}-M_{N}$ with $E_{N}$ being the single-particle energy.
With the nucleon self-energies shown in Eqs.~\eqref{eq:self-engy-scalar}--\eqref{eq:self-engy-space}, the single-particle energy is given by a solution of the transcendental equation,
\begin{equation}
  E_{N}(k) = \left[E_{N}^{\ast}(k)-\Sigma_{N}^{0}(k)\right]_{k^{0}=E_{N}(k)}.
  \label{eq:transcendental-equation}
\end{equation}
%

%%%%%%%%%%%%%%%%%%%%%%%%%%%%%%%%%%%%%%%%%%%%%%%%%%%%%%%%%%%%%%%%%%%%%%%%%%%%%%%%
\begin{figure}%[h!]
\includegraphics[width=11.0cm,keepaspectratio,clip]{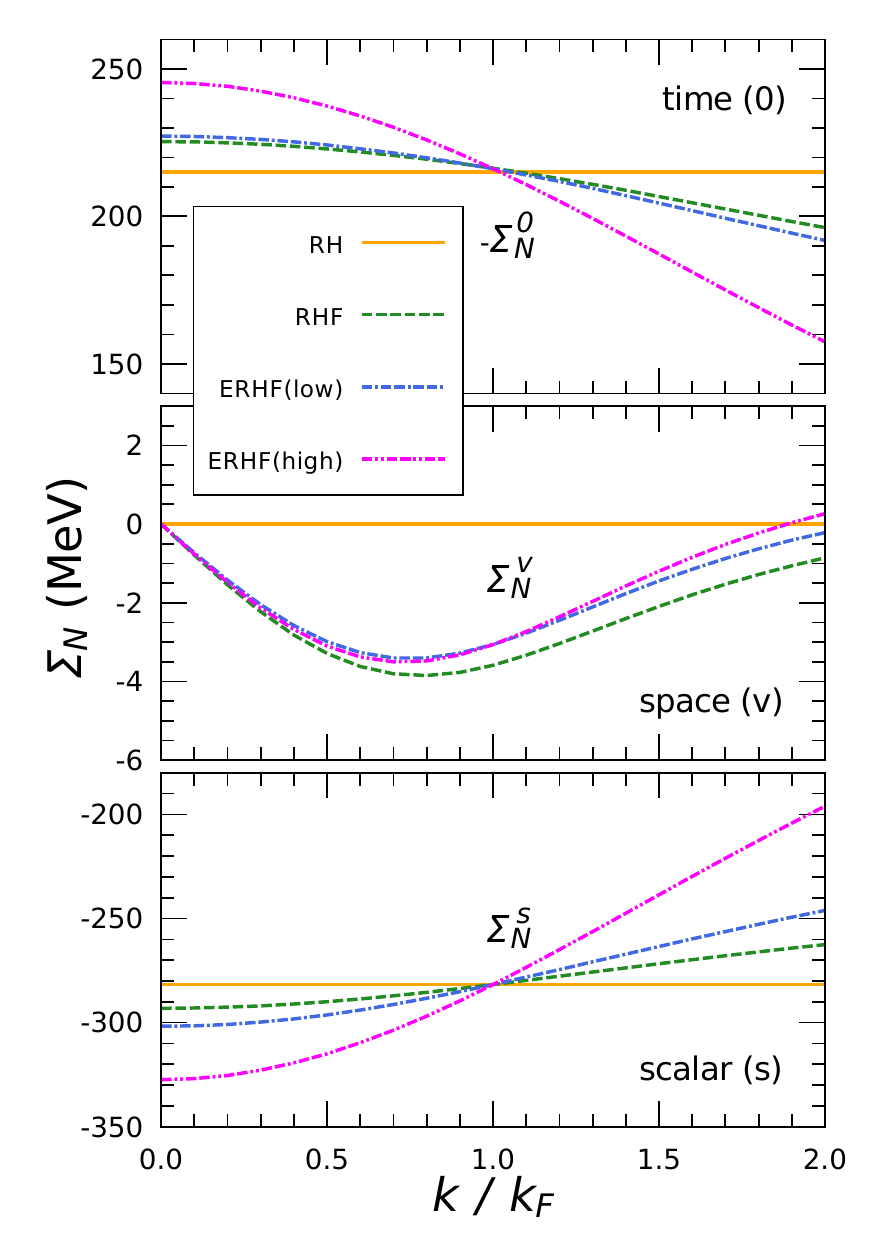}
\caption{\label{fig:NSE-momentum}
Momentum dependence of nucleon self-energies in symmetric nuclear matter at $\rho_{0}$.
The top (middle) [bottom] panel is for the time (space) [scalar] component, $\Sigma_{N}^{0(v)[s]}$.
}
\end{figure}
%%%%%%%%%%%%%%%%%%%%%%%%%%%%%%%%%%%%%%%%%%%%%%%%%%%%%%%%%%%%%%%%%%%%%%%%%%%%%%%%
The momentum dependence of nucleon self-energies, $\Sigma_{N}^{s,0,v}$, in symmetric nuclear matter at $\rho_{0}$ is shown in Figs.~\ref{fig:NSE-momentum}.
We present the results of the RH, RHF, and ERHF models, which satisfy the saturation conditions and $E_{\rm sym}$ at $\rho_{0}$.
Because the direct contribution of $\Sigma_{N}^{s,0,v}$ directly couples to the mean-field values of the $\sigma$, $\omega$, and $\rho$ mesons, all the components retain the constant values at any momentum in the RH model.
It is found that the $\Sigma_{N}^{v}$ does not show any impact in symmetric nuclear matter within Hartree approximation, and that, even in RHF approximation, it is very small.
In contrast, the momentum dependence due to the exchange contribution is clearly demonstrated in the RHF and ERHF models.
Thus, it is expected that the self-energies, $\Sigma_{N}^{s}$ and $\Sigma_{N}^{0}$, contribute dominantly to the single-nucleon potential at $\rho_{0}$.

%%%%%%%%%%%%%%%%%%%%%%%%%%%%%%%%%%%%%%%%%%%%%%%%%%%%%%%%%%%%%%%%%%%%%%%%%%%%%%%%
\begin{figure}
\includegraphics[width=11.0cm,keepaspectratio,clip]{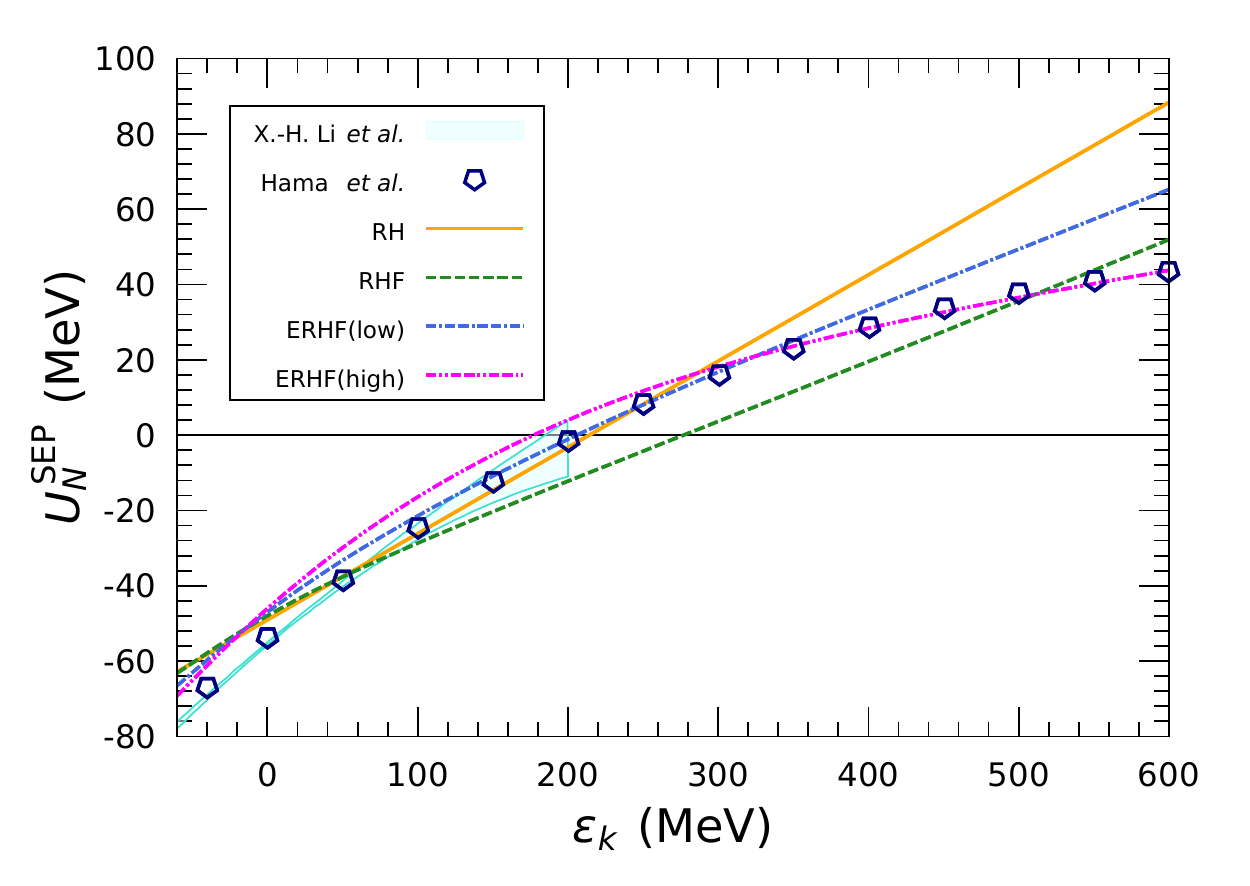}
\caption{\label{fig:USEP}
Energy dependence of single-nucleon potential, $U_{N}^{\rm SEP}$, in symmetric nuclear matter at $\rho_{0}$.
The shaded band shows the result of the nucleon-optical-model potential extracted from analyzing the nucleon-nucleus scattering data~\cite{Li:2013ck}, denoted by X.-H.~Li {\it et al.}
The results of the Schr\"{o}dinger-equivalent potential obtained by the Dirac phenomenology for elastic proton-nucleus scattering data calculated by Hama {\it et al.}~\cite{Hama:1990vr} are also included.
}
\end{figure}
%%%%%%%%%%%%%%%%%%%%%%%%%%%%%%%%%%%%%%%%%%%%%%%%%%%%%%%%%%%%%%%%%%%%%%%%%%%%%%%%
The energy dependence of single-nucleon potential (or nucleon optical potential), $U_{N}^{\rm SEP}$, is depicted in Fig.~\ref{fig:USEP}.
We also show the results of the RH, RHF and ERHF models.
In the ERHF low (high) model, the coupling constants, $\tilde{g}_{\sigma}$, $\tilde{g}_{\omega}$, and $\tilde{g}_{\rho}$, are adjusted so as to cover the scattering data for $\epsilon_{k} < (>) \, 300$ MeV.
As $\Sigma_{N}^{s}$ and $\Sigma_{N}^{0}$ are constant and $U_{N}^{\rm SEP}$ is proportional to $\epsilon_{k}$ in the RH model, it is difficult to reproduce the scattering data widely in RH approximation.
Meanwhile, due to the momentum dependence which is intrinsically possessed in $\Sigma_{N}^{s,0,v}$ through Fock terms, $U_{N}^{\rm SEP}$ depends on $\epsilon_{k}$ non-linearly in the RHF and ERHF models.
Moreover, it is found that, in the ERHF(high) model, the enhanced exchange contribution makes it possible to well reproduce the scattering data at high $\epsilon_{k}$.
We note that $U_{N}^{\rm SEP}$ strongly depends on the effective nucleon mass~\cite{Danielewicz:1999zn},  which is fixed as $M_{N}^{\ast}/M_{N}=0.70$ at $\rho_{0}$ in the present calculation.

%%%%%%%%%%%%%%%%%%%%%%%%%%%%%%%%%%%%%%%%%%%%%%%%%%%%%%%%%%%%%%%%%%%%%%%%%%%%%%%%
%% Table: Self-energy
%%%%%%%%%%%%%%%%%%%%%%%%%%%%%%%%%%%%%%%%%%%%%%%%%%%%%%%%%%%%%%%%%%%%%%%%%%%%%%%%
\begin{table*}
\caption{\label{tab:self-energy}
Contents of nucleon self-energies, $\Sigma_{N}^{s,0,v}$, in symmetric nuclear matter at $\rho_{0}$.
The values are in MeV.
}
\begin{ruledtabular}
\begin{tabular}{llrrrrrrrrrrrr}
\                &                 \ &
\multicolumn{3}{c}{RH}                                                 & \multicolumn{3}{c}{RHF} &
\multicolumn{3}{c}{ERHF(low)}                                          & \multicolumn{3}{c}{ERHF(high)}                                          \\
\cline{3-5}\cline{6-8}\cline{9-11}\cline{12-14}
\                &                 \ &
$\Sigma_{N}^{s}$ & $-\Sigma_{N}^{0}$ & \hspace{0.15cm} $\Sigma_{N}^{v}$ & $\Sigma_{N}^{s}$ & $-\Sigma_{N}^{0}$ & \hspace{0.15cm} $\Sigma_{N}^{v}$ &
$\Sigma_{N}^{s}$ & $-\Sigma_{N}^{0}$ & \hspace{0.15cm} $\Sigma_{N}^{v}$ & $\Sigma_{N}^{s}$ & $-\Sigma_{N}^{0}$ & \hspace{0.15cm} $\Sigma_{N}^{v}$ \\
\hline
Direct           &          $\sigma$ &
          $-282$ &               $0$ &                              $0$ &           $-240$ &              $0$ &                              $0$ &
          $-201$ &               $0$ &                              $0$ &            $-83$ &              $0$ &                              $0$ \\
\                &          $\omega$ &
             $0$ &             $215$ &                              $0$ &              $0$ &            $165$ &                              $0$ &
             $0$ &             $155$ &                              $0$ &              $0$ &             $74$ &                              $0$ \\
Exchange         &          $\sigma$ &
              -- &                -- &                               -- &             $26$ &             $27$ &                             $-1$ &
            $22$ &              $23$ &                             $-1$ &             $41$ &            $109$ &                             $-2$ \\
\                &          $\omega$ &
              -- &                -- &                               -- &            $-57$ &             $30$ &                             $-1$ &
           $-86$ &              $45$ &                             $-1$ &           $-209$ &             $21$ &                             $-3$ \\
\                &             $\pi$ &
              -- &                -- &                               -- &             $-4$ &             $-4$ &                             $-3$ &
            $-4$ &              $-4$ &                             $-3$ &             $-4$ &             $-4$ &                             $-3$ \\
\                &       $\rho_{VV}$ &
              -- &                -- &                               -- &             $-1$ &              $1$ &                              $0$ &
            $-2$ &               $1$ &                              $0$ &             $-5$ &              $3$ &                              $0$ \\
\                &       $\rho_{TT}$ &
              -- &                -- &                               -- &             $-6$ &             $-2$ &                              $0$ &
           $-10$ &              $-4$ &                              $0$ &            $-23$ &             $-8$ &                             $-1$ \\
\                &       $\rho_{VT}$ &
              -- &                -- &                               -- &              $0$ &              $0$ &                              $2$ &
             $1$ &               $0$ &                              $2$ &              $1$ &              $0$ &                              $6$ \\
Total            &                 \ &
          $-282$ &             $215$ &                              $0$ &           $-282$ &            $216$ &                             $-4$ &
          $-282$ &             $216$ &                             $-3$ &           $-282$ &            $216$ &                             $-3$ \\
\end{tabular}
\end{ruledtabular}
\end{table*}
%%%%%%%%%%%%%%%%%%%%%%%%%%%%%%%%%%%%%%%%%%%%%%%%%%%%%%%%%%%%%%%%%%%%%%%%%%%%%%%%
The contents of nucleon self-energies, $\Sigma_{N}^{s,0,v}$, in symmetric nuclear matter at $\rho_{0}$ are presented in Table~\ref{tab:self-energy}.
In the RH model, an attractive (repulsive) force comes from only the direct contribution due to the $\sigma$ ($\omega$) meson exchange through $\Sigma_{N}^{s}$ ($-\Sigma_{N}^{0}$).
On the other hand, in RHF approximation, all the components of $\Sigma_{N}$ are affected by the exchange contribution.
As for the exchange contribution in the RHF and ERHF models, the $\sigma$ ($\omega$) meson gives a repulsive (attractive) force in the scalar component, while both $\sigma$ and $\omega$ mesons work as a repulsive force in the time component.
Although the pion also influences all the components through the exchange diagram, its contribution is small.
Moreover, even in symmetric nuclear matter, the $\rho$ meson contributes to $\Sigma_{N}$ through Fock terms, where the contribution due to tensor-tensor ($TT$) mixing is relatively large comparing with those due to vector-vector ($VV$) and vector-tensor ($VT$) mixing.
We note that $\Sigma_{N}^{v}$ is very small at $\rho_{0}$.

%%%%%%%%%%%%%%%%%%%%%%%%%%%%%%%%%%%%%%%%%%%%%%%%%%%%%%%%%%%%%%%%%%%%%%%%%%%%%%%%
\begin{figure}
\includegraphics[width=11.0cm,keepaspectratio,clip]{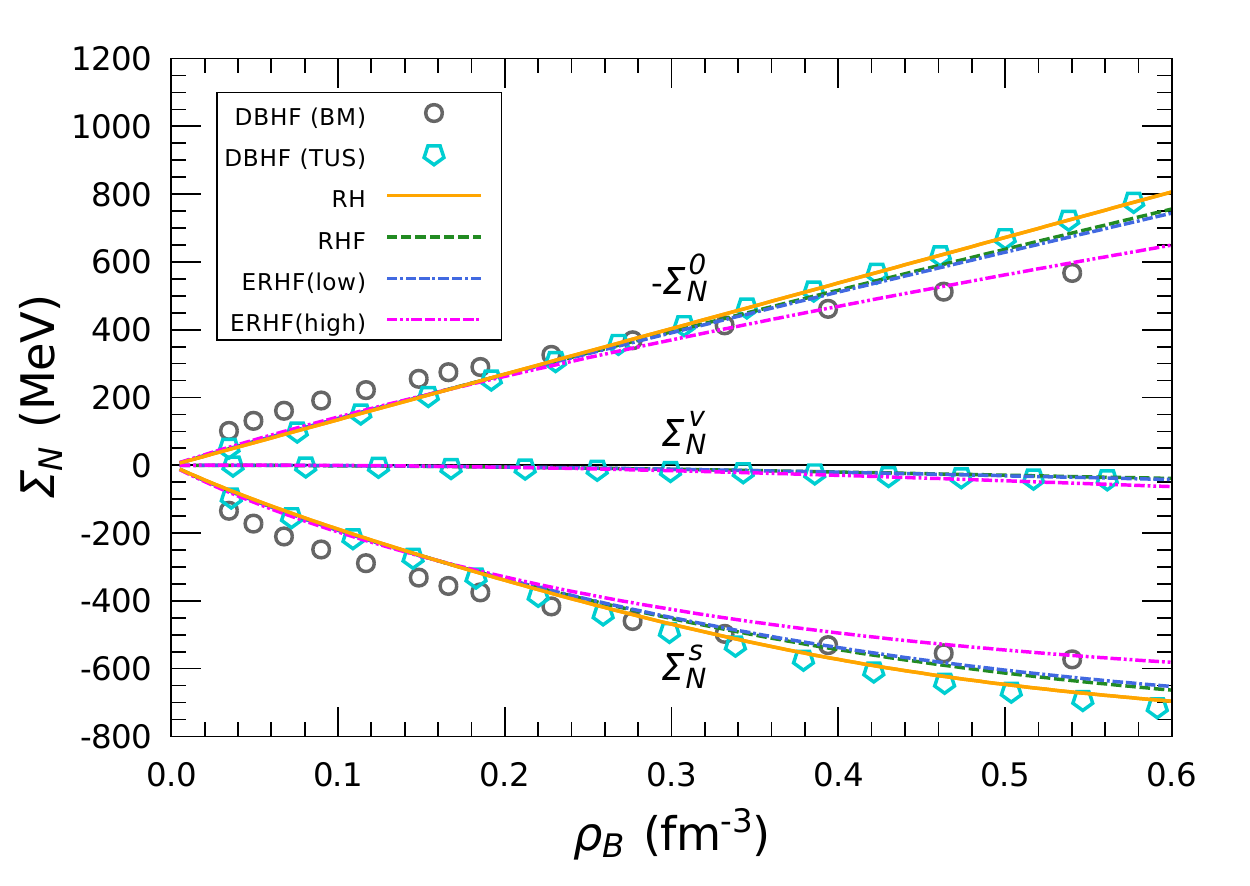}
\caption{\label{fig:NSE-density}
Nucleon self-energies, $\Sigma_{N}^{s,0,v}$, in symmetric nuclear matter as a function of $\rho_{B}$.
The $\Sigma_{N}^{s,0,v}$ are given at the Fermi surface, $k_{F}$.
The DBHF results %with the Bonn A potential denoted
by Brockmann and Machleidt (BM)~\cite{Brockmann:1990cn} and Katayama and Saito  (TUS)~\cite{Katayama:2013zya} are also presented.
}
\end{figure}
%%%%%%%%%%%%%%%%%%%%%%%%%%%%%%%%%%%%%%%%%%%%%%%%%%%%%%%%%%%%%%%%%%%%%%%%%%%%%%%%
\begin{figure}[t!]
\includegraphics[width=11.0cm,keepaspectratio,clip]{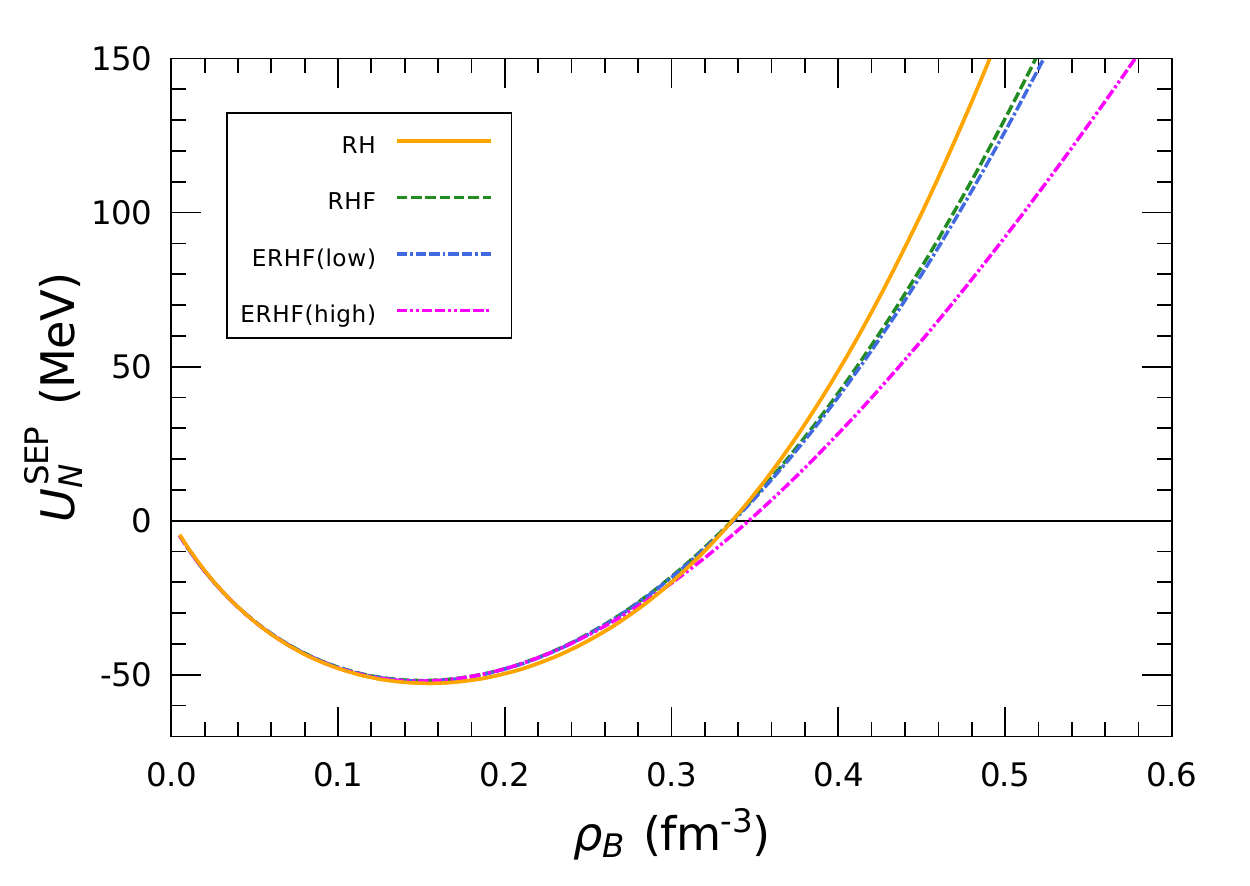}
\caption{\label{fig:USEP-density}
% Schr\"{o}dinger-equivalent potential, $U_{N}^{\rm SEP}$, at $k_{F}$ in symmetric nuclear matter as a function of $\rho_{B}$.
Single-nucleon potential, $U_{N}^{\rm SEP}$, at $k_{F}$ in symmetric nuclear matter as a function of $\rho_{B}$.
}
\end{figure}
%%%%%%%%%%%%%%%%%%%%%%%%%%%%%%%%%%%%%%%%%%%%%%%%%%%%%%%%%%%%%%%%%%%%%%%%%%%%%%%%
In Figs.~\ref{fig:NSE-density} and \ref{fig:USEP-density}, the nucleon self-energies, $\Sigma_{N}^{s,0,v}$, and the single-nucleon potential, $U_{N}^{\rm SEP}$, in symmetric nuclear matter are respectively presented as a function of $\rho_{B}$.
In the present calculation, the model dependence of $\Sigma_{N}^{s,0,v}$ and $U_{N}^{\rm SEP}$ is very weak at densities below $\rho_{0}$, while the Fock terms play important roles in both $\Sigma_{N}^{s,0,v}$ and $U_{N}^{\rm SEP}$ at high densities.
It is also interesting to compare the density dependence of $\Sigma_{N}^{s,0,v}$ with the Dirac-Brueckner-Hartree-Fock (DBHF) calculation.
The self-energies, $\Sigma_{N}^{s}$ and $-\Sigma_{N}^{0}$, in RH approximation, are very similar to those in the DBHF result by the TUS group~\cite{Katayama:2013zya}, whereas, with increasing $\rho_{B}$, $\Sigma_{N}^{s}$ and $-\Sigma_{N}^{0}$ in the ERHF(high) model become close to the results calculated by Brockmann and Machleidt~\cite{Brockmann:1990cn}.
In Fig.~\ref{fig:USEP-density}, although any Fock effect on $U_{N}^{\rm SEP}$ is little seen up to $2\rho_{0}$, the exchange terms give a large contribution to $U_{N}^{\rm SEP}$ at high densities.

%%%%%%%%%%%%%%%%%%%%%%%%%%%%%%%%%%%%%%%%%%%%%%%%%%%%%%%%%%%%%%%%%%%%%%%%%%%%%%%%
\subsection{Effective nucleon mass and nuclear equation of state}
\label{subsec:nuclear-matter}
%%%%%%%%%%%%%%%%%%%%%%%%%%%%%%%%%%%%%%%%%%%%%%%%%%%%%%%%%%%%%%%%%%%%%%%%%%%%%%%%

%%%%%%%%%%%%%%%%%%%%%%%%%%%%%%%%%%%%%%%%%%%%%%%%%%%%%%%%%%%%%%%%%%%%%%%%%%%%%%%%
\begin{figure}
\includegraphics[width=11.0cm,keepaspectratio,clip]{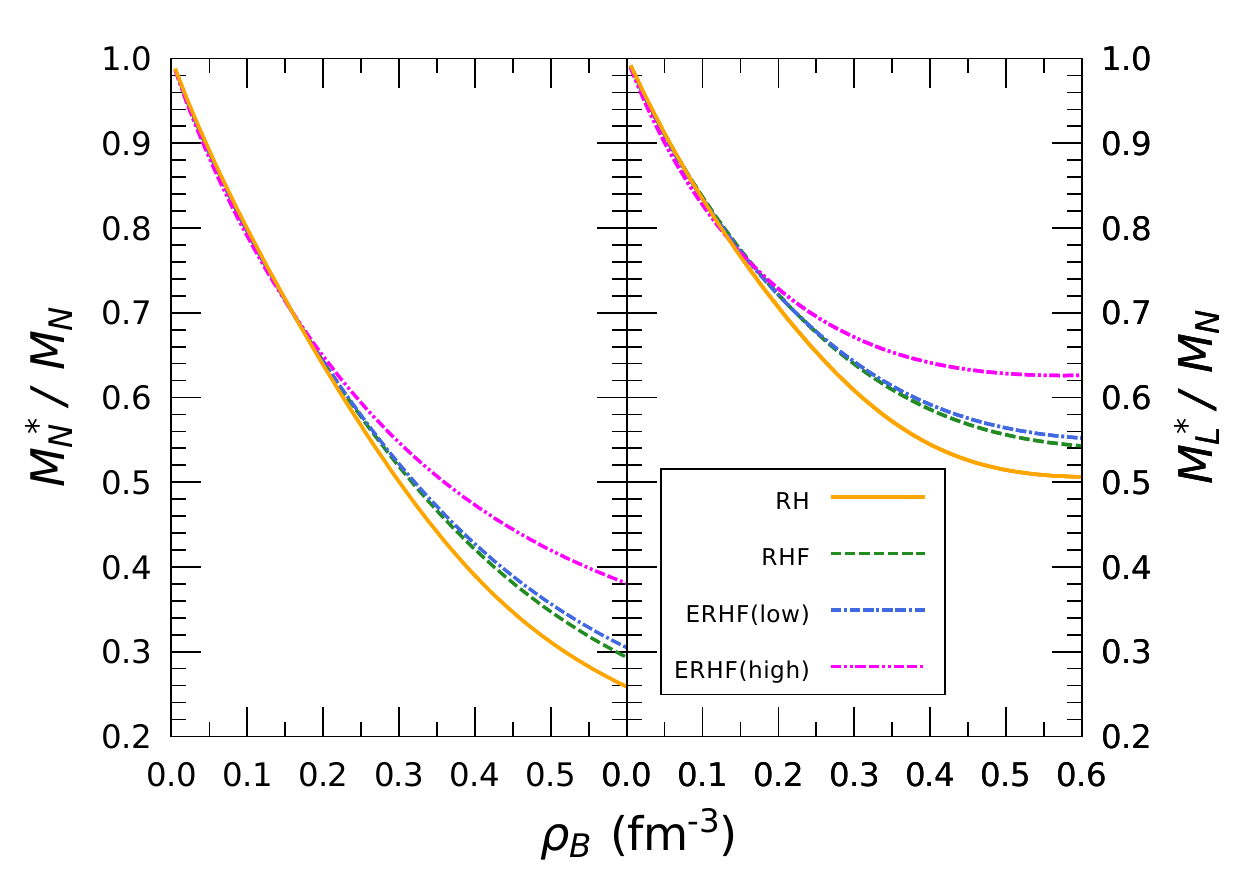}%
\caption{\label{fig:Emass}
Effective nucleon masses in symmetric nuclear matter as a function of $\rho_{B}$.
The left panel is for the calculation of the relativistic mass, $M_{N}^{\ast}$, and the right one is for that of the Landau (nonrelativistic) mass, $M_{L}^{\ast}$.
}
\end{figure}
%%%%%%%%%%%%%%%%%%%%%%%%%%%%%%%%%%%%%%%%%%%%%%%%%%%%%%%%%%%%%%%%%%%%%%%%%%%%%%%%
The density dependence of the effective nucleon mass in symmetric nuclear matter is displayed in Fig.~\ref{fig:Emass}.
We show two types of the effective nucleon mass: one is the relativistic mass in matter, $M_{N}^{\ast}$, defined in Eq.~\eqref{eq:auxiliary-quantity-mass}, and the other is the effective mass in a nonrelativistic framework, $M_{L}^{\ast}$, which is the so-called Landau mass~\cite{Maruyama:1999ye,Typel:2005ba}.
Compared with the Landau mass, the relativistic one decreases rapidly as the density increases.
We can see that both masses in the RH model are smaller than those in the other models at the densities above $\rho_{0}$.
It is also found that the exchange contribution suppresses their sharp reduction at high densities.

%%%%%%%%%%%%%%%%%%%%%%%%%%%%%%%%%%%%%%%%%%%%%%%%%%%%%%%%%%%%%%%%%%%%%%%%%%%%%%%%
\begin{figure}
\includegraphics[width=11.0cm,keepaspectratio,clip]{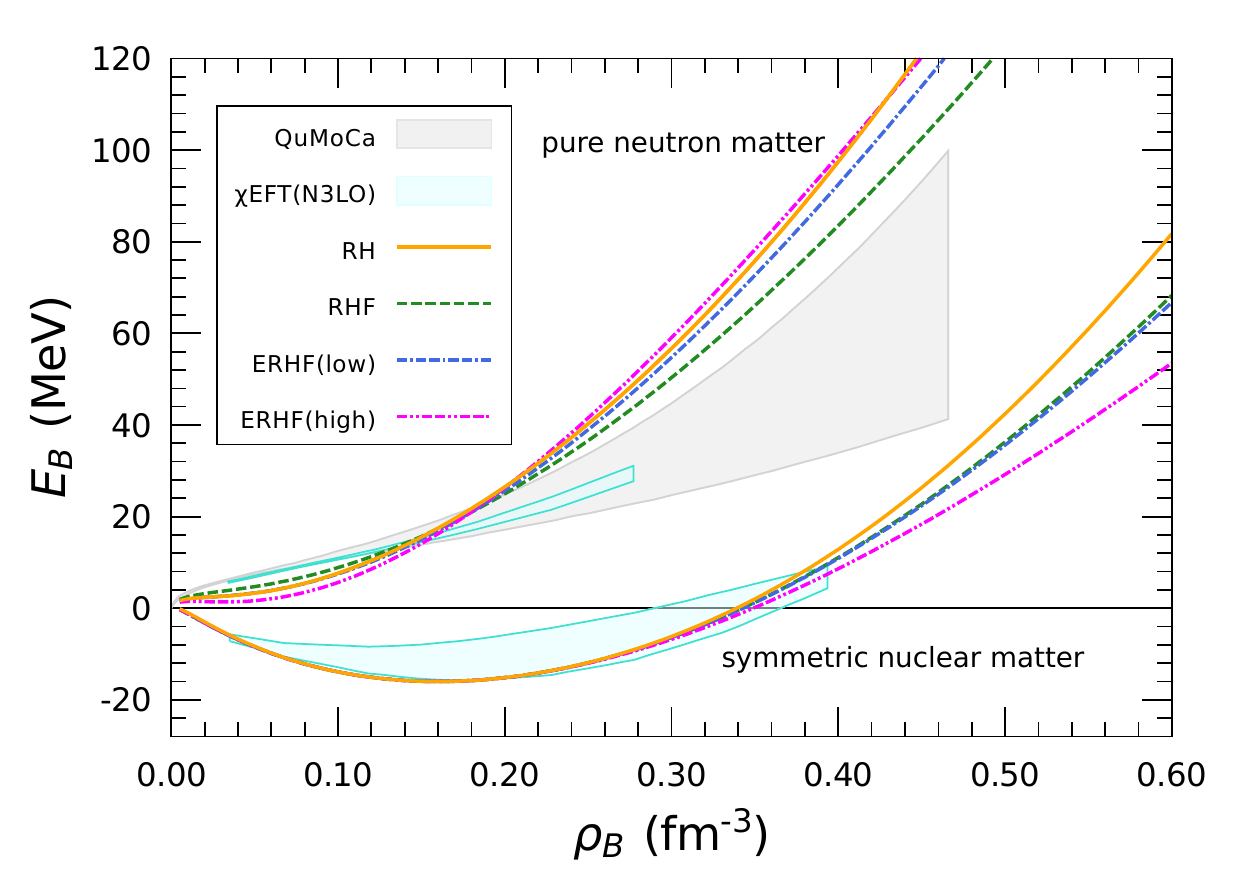}%
\caption{\label{fig:Binding-energy}
Nuclear binding energy per nucleon, $E_{B}$, for symmetric nuclear or pure neutron matter as a function of $\rho_{B}$.
For comparison, the results based on $\chi$EFT with N$^{3}$LO two-nucleon and N$^{2}$LO three-nucleon forces~\cite{Sammarruca:2014zia} and the QuMoCa method~\cite{Gandolfi:2011xu} are also presented.
}
\end{figure}
%%%%%%%%%%%%%%%%%%%%%%%%%%%%%%%%%%%%%%%%%%%%%%%%%%%%%%%%%%%%%%%%%%%%%%%%%%%%%%%%
In Fig.~\ref{fig:Binding-energy}, the nuclear binding energy per nucleon, $E_{B}$, for symmetric nuclear or pure neutron matter is presented.
For symmetric nuclear matter, our results are close to the recent calculation based on  chiral effective field theory ($\chi$EFT) in the density region below 0.4 fm$^{-3}$~\cite{Sammarruca:2014zia}.
We see that the Fock contribution diminishes $E_{B}$ at high densities.
In contrast, for pure neutron matter, the effect of Fock terms is not small even at low densities.
Moreover, the present results tend to be larger than those calculated by $\chi$EFT and Quantum Monte Carlo (QuMoCa) method at high densities~\cite{Gandolfi:2011xu}.
It is found that, as seen in the ERHF(high) model, the Fock terms enhances the difference between $E_{B}$ for symmetric nuclear and pure neutron matter, which implies that a  large exchange contribution enlarges $E_{\rm sym}$ at the densities above $\rho_{0}$.

%%%%%%%%%%%%%%%%%%%%%%%%%%%%%%%%%%%%%%%%%%%%%%%%%%%%%%%%%%%%%%%%%%%%%%%%%%%%%%%%
\begin{figure}
\includegraphics[width=11.0cm,keepaspectratio,clip]{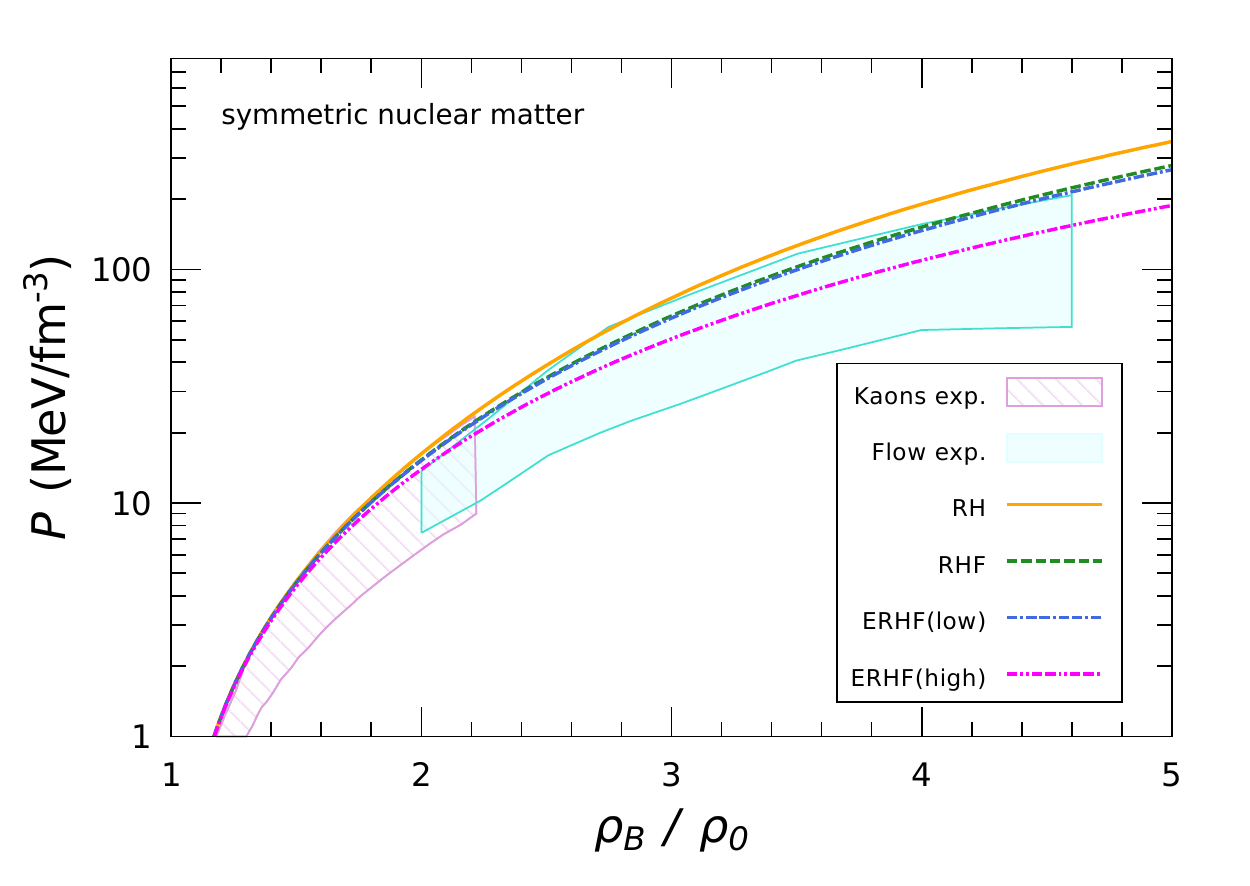}%
\\
\includegraphics[width=11.0cm,keepaspectratio,clip]{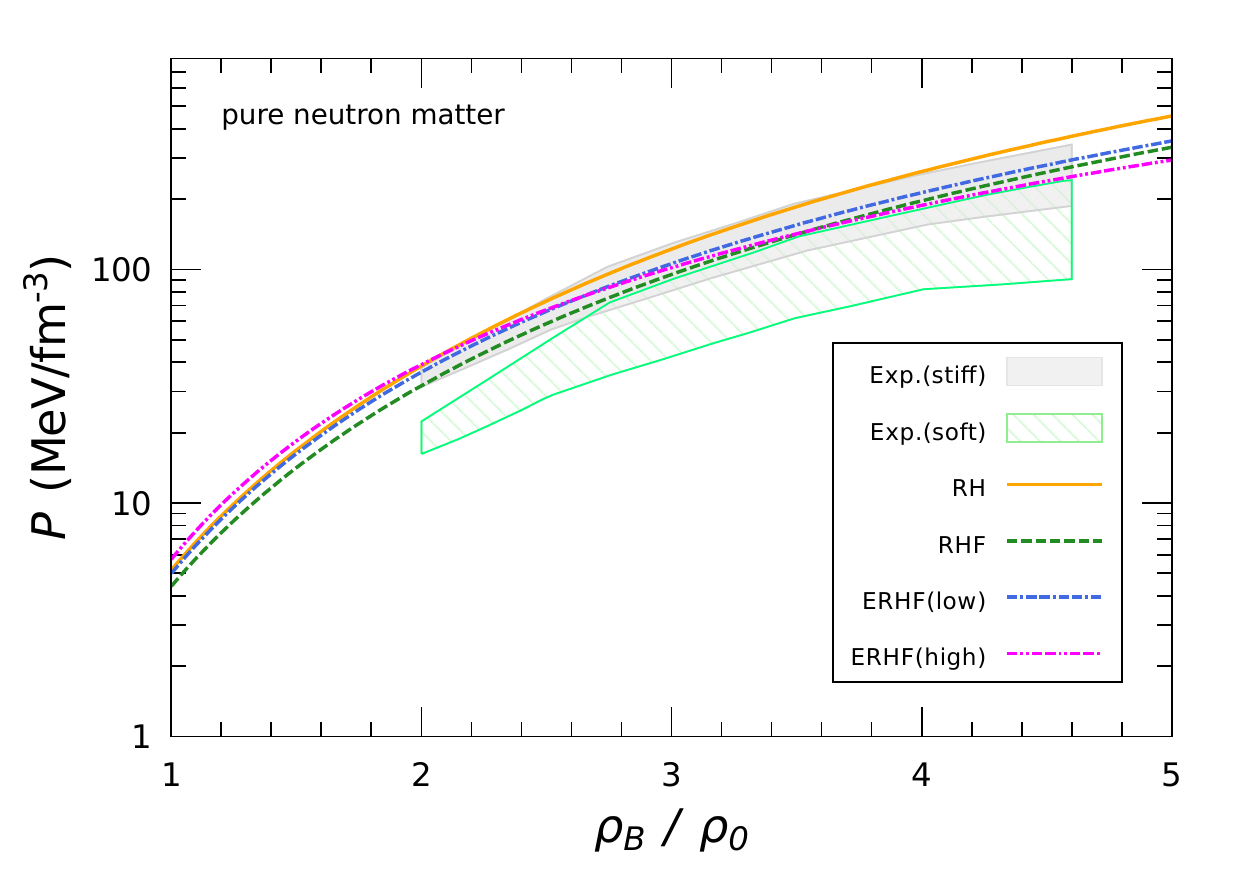}%
\caption{\label{fig:Pressure}
Pressure for symmetric nuclear or pure neutron matter, $P$, as a function of the density ratio, $\rho_{B}/\rho_{0}$.
The upper (lower) panel is for the case of symmetric nuclear (pure neutron) matter.
The experimental constraints on the nuclear equation of state from heavy-ion flow data are presented~\cite{Danielewicz:2002pu,Fuchs:2005zg}.
For pure neutron matter, the flow data is estimated with stiff or soft density dependence.
}
\end{figure}
%%%%%%%%%%%%%%%%%%%%%%%%%%%%%%%%%%%%%%%%%%%%%%%%%%%%%%%%%%%%%%%%%%%%%%%%%%%%%%%%
In Fig.~\ref{fig:Pressure}, we illustrate pressure for symmetric nuclear or pure neutron matter in comparison with the experimental constraints from heavy-ion flow data~\cite{Danielewicz:2002pu,Fuchs:2005zg}.
In both cases, pressure in the RH model exceeds the constraints at high densities, while those in the RHF and ERHF models are consistent with the analysis of heavy-ion collision data.
It is found that the exchange contribution softens pressure at high densities.
% In contrast, as seen in the ERHF(high) model, the large Fock contribution enhances pressure in pure neutron matter at low densities.

%%%%%%%%%%%%%%%%%%%%%%%%%%%%%%%%%%%%%%%%%%%%%%%%%%%%%%%%%%%%%%%%%%%%%%%%%%%%%%%%
\subsection{Symmetry energy and its slope parameter}
\label{subsec:Esym-L}
%%%%%%%%%%%%%%%%%%%%%%%%%%%%%%%%%%%%%%%%%%%%%%%%%%%%%%%%%%%%%%%%%%%%%%%%%%%%%%%

%%%%%%%%%%%%%%%%%%%%%%%%%%%%%%%%%%%%%%%%%%%%%%%%%%%%%%%%%%%%%%%%%%%%%%%%%%%%%%%%
%% Table: Matter-properties
%%%%%%%%%%%%%%%%%%%%%%%%%%%%%%%%%%%%%%%%%%%%%%%%%%%%%%%%%%%%%%%%%%%%%%%%%%%%%%%%
\begin{table*}%[h!]
\caption{\label{tab:matter-properties}
Properties of symmetric nuclear matter at $\rho_{0}$.
The incompressibility and effective nucleon mass are respectively fixed as $K_{0}=250$ MeV and $M_{N}^{\ast}/M_{N}=0.70$.
The nuclear symmetry energy in the models without asterisk is also fitted so as to reproduce the empirical data, $E_{\rm sym}(\rho_{0})=32.5$ MeV.
The physical quantities are explained for details in the text.
}
\begin{ruledtabular}
\begin{tabular}{lcccccccc}
\            & $M_{L}^{\ast}/M_{N}$ & $J_{0}$ & $E_{\rm sym}$ & $L$   & $K_{\rm sym}$ & $K_{\rm asy}$ & $K_{{\rm sat},2}$ \\
Model         &                   \ &   (MeV) &         (MeV) & (MeV) &         (MeV) &         (MeV) &             (MeV) \\
\hline
RH$^{\ast}$  &                0.754 &  $-361$ &          23.5 &  67.0 &          32.0 &        $-370$ &            $-273$ \\
RH           &                0.754 &  $-361$ &          32.5 &  94.1 &          32.0 &        $-533$ &            $-397$ \\
RHF$^{\ast}$ &                0.733 &  $-368$ &          46.3 & 123.6 &       $-41.0$ &        $-783$ &            $-601$ \\
RHF          &                0.763 &  $-353$ &          32.5 &  81.1 &       $-17.5$ &        $-504$ &            $-390$ \\
ERHF(low)    &                0.762 &  $-350$ &          32.5 &  94.6 &          42.2 &        $-526$ &            $-393$ \\
ERHF(high)   &                0.762 &  $-417$ &          32.5 & 113.5 &         116.6 &        $-564$ &            $-375$ \\
\end{tabular}
\end{ruledtabular}
\end{table*}
%%%%%%%%%%%%%%%%%%%%%%%%%%%%%%%%%%%%%%%%%%%%%%%%%%%%%%%%%%%%%%%%%%%%%%%%%%%%%%%%
The properties of symmetric nuclear matter at $\rho_{0}$ is presented in Table~\ref{tab:matter-properties}.
The Landau mass of nucleon and the third-order incompressibility are denoted by $M_{L}^{\ast}$ and $J_{0}$, respectively.
The nuclear symmetry energy, $E_{\rm sym}$, around $\rho_{0}$ is approximately expressed as a power series of the isospin-asymmetry parameter, $\delta=\left(\rho_{n}-\rho_{p}\right)/\rho_{B}$~\cite{Chen:2007ih,Chen:2009wv}:
\begin{align}
  E_{\rm sym}(\rho_{B})
  &= \left.\frac{1}{2!}\frac{\partial^{2}E_{B}(\rho_{B},\delta)}{\partial\delta^{2}}\right|_{\delta=0}
  \nonumber \\
  &\simeq E_{\rm sym}(\rho_{0}) + L(\rho_{0})\left(\frac{\rho_{B}-\rho_{0}}{3\rho_{0}}\right)
  +       \frac{K_{\rm sym}(\rho_{0})}{2!}\left(\frac{\rho_{B}-\rho_{0}}{3\rho_{0}}\right)^{2},
\end{align}
with
\begin{align}
  L(\rho_{0})           &= 3\rho_{0}\left.\frac{dE_{\rm sym}(\rho_{B})}{d\rho_{B}}\right|_{\rho_{B}=\rho_{0}},
  \\
  K_{\rm sym}(\rho_{0}) &= 9\rho_{0}^{2}\left.\frac{d^{2}E_{\rm sym}(\rho_{B})}{d\rho_{B}^{2}}\right|_{\rho_{B}=\rho_{0}},
\end{align}
where $L$ and $K_{\rm sym}$ are respectively the slope and curvature parameters.
In addition, using this parabolic approximation, the 2nd derivative of the isobaric incompressibility coefficient is given by $K_{{\rm sat},2}=K_{\rm asy}-\frac{J_{0}}{K_{0}}L$ with the parameter $K_{\rm asy}=K_{\rm sym}-6L$.
Although $K_{{\rm sat},2}$ is model dependent~\cite{Dutra:2014qga,Stone:2014wza}, we see that the present calculations in the RH, RHF and ERHF models provide $K_{{\rm sat},2}=-390\pm15$ MeV, which is consistent with the empirical constraints, $K_{{\rm sat},2}=-370\pm120$ MeV~\cite{Chen:2009wv}.
This fact may imply that $K_{{\rm sat},2}$ is not much affected by the exchange contribution around $\rho_{0}$.

%%%%%%%%%%%%%%%%%%%%%%%%%%%%%%%%%%%%%%%%%%%%%%%%%%%%%%%%%%%%%%%%%%%%%%%%%%%%%%%%
\begin{figure}% [h!]
\includegraphics[width=11.0cm,keepaspectratio,clip]{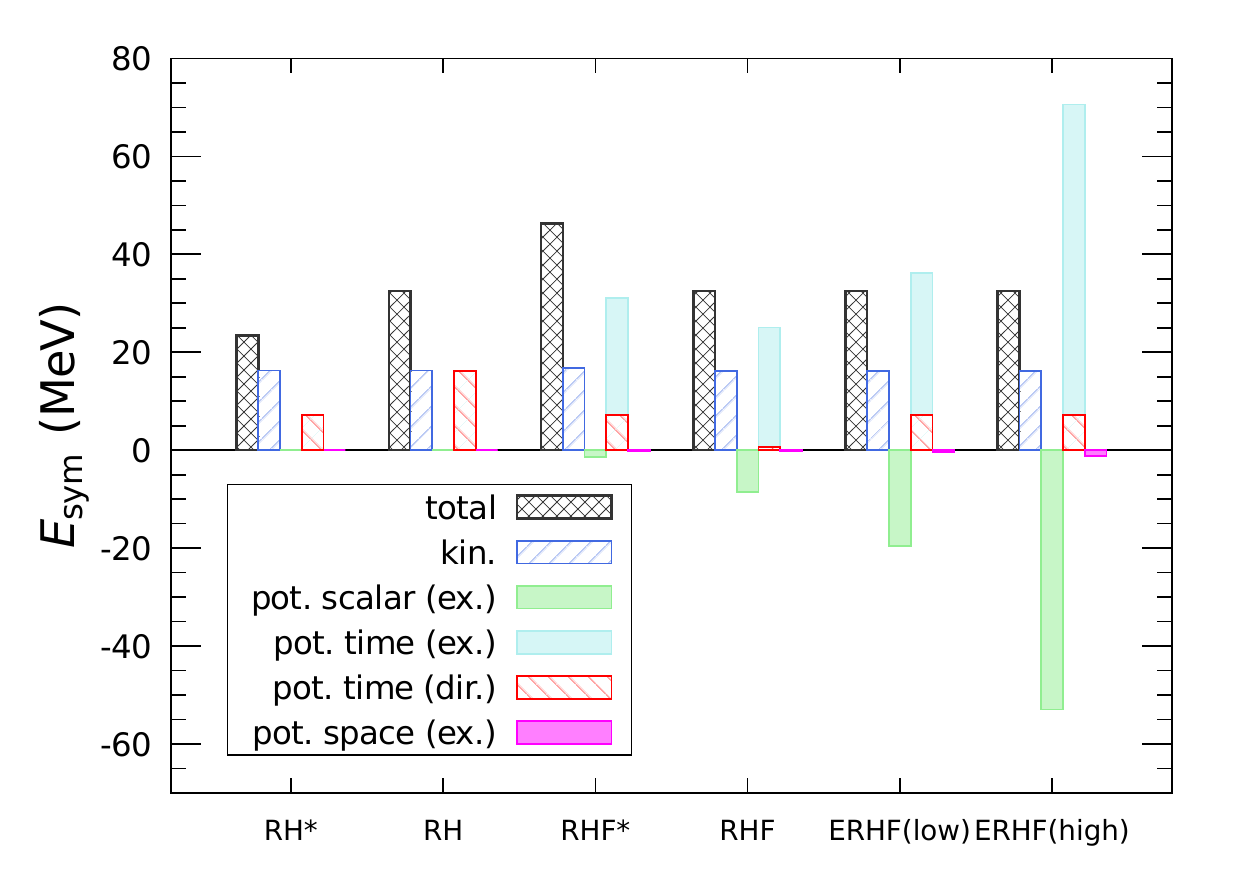}
\caption{\label{fig:Esym-rho0}
Components of $E_{\rm sym}$ at $\rho_{0}$. See the text for details.
}
\end{figure}
%%%%%%%%%%%%%%%%%%%%%%%%%%%%%%%%%%%%%%%%%%%%%%%%%%%%%%%%%%%%%%%%%%%%%%%%%%%%%%%%
In Fig.~\ref{fig:Esym-rho0}, we present the detail of nuclear symmetry energy, $E_{\rm sym}$, at $\rho_{0}$.
As shown in Eq.~\eqref{eq:Esym}, $E_{\rm sym}$ can be separated into the kinetic and potential terms, $E_{\rm sym}^{\rm kin}$ and $E_{\rm sym}^{\rm pot}$.
In addition, $E_{\rm sym}^{\rm pot}$ is divided into the scalar, time and space components, based on the Lorentz structure of nucleon self-energies.
The direct contribution in the potential term, $E_{\rm sym}^{\rm pot, dir}$, only comes from the time component via the $\rho$ meson shown in Eq.~\eqref{eq:Esym-pot-dir}.

It is found that $E_{\rm sym}^{\rm kin}$ is approximately 16 MeV in all the models because it is primarily determined by the $M_{N}^{\ast}$ and $k_{F_{N}}^{\ast}$ at $\rho_{0}$.
In the RH$^{\ast}$ model, because of the inadequate contribution of $E_{\rm sym}^{\rm pot}$ ($=E_{\rm sym}^{\rm pot, dir}$), $E_{\rm sym}$ cannot reach the experimental value, as already shown in Table~\ref{tab:matter-properties}.
In contrast, in the RHF$^{\ast}$ model, $E_{\rm sym}$ is overestimated since the exchange contribution due to the time component enlarges $E_{\rm sym}^{\rm pot}$ considerably.
On the other hand, in the RH, RHF and ERHF models, $E_{\rm sym}$ is fitted to be the empirical value, 32.5 MeV, by changing the $\rho$-$N$ coupling constant or by  enhancing the exchange contributions with the new coupling constants, $w_M$.
As a result, the $E_{\rm sym}^{\rm pot, dir}$ in the RH model is larger than that in the RH$^{\ast}$ model, while that in the RHF model becomes quite small.
In the ERHF models, because we use the empirical value, $g_{\rho}^{2}/4\pi=0.55$, the direct contribution is constant, $E_{\rm sym}^{\rm pot,dir}=7.2$ MeV, in both low and high cases.
Therefore, the exchange contribution in the potential term, $E_{\rm sym}^{\rm pot, ex}$, is totally estimated to be around 9.3 MeV, which is mainly composed by a cancellation of large positive and negative values due to the time and scalar components, respectively.
The space component is negligible at $\rho_{0}$.

%%%%%%%%%%%%%%%%%%%%%%%%%%%%%%%%%%%%%%%%%%%%%%%%%%%%%%%%%%%%%%%%%%%%%%%%%%%%%%%%
\begin{figure}
\includegraphics[width=11.0cm,keepaspectratio,clip]{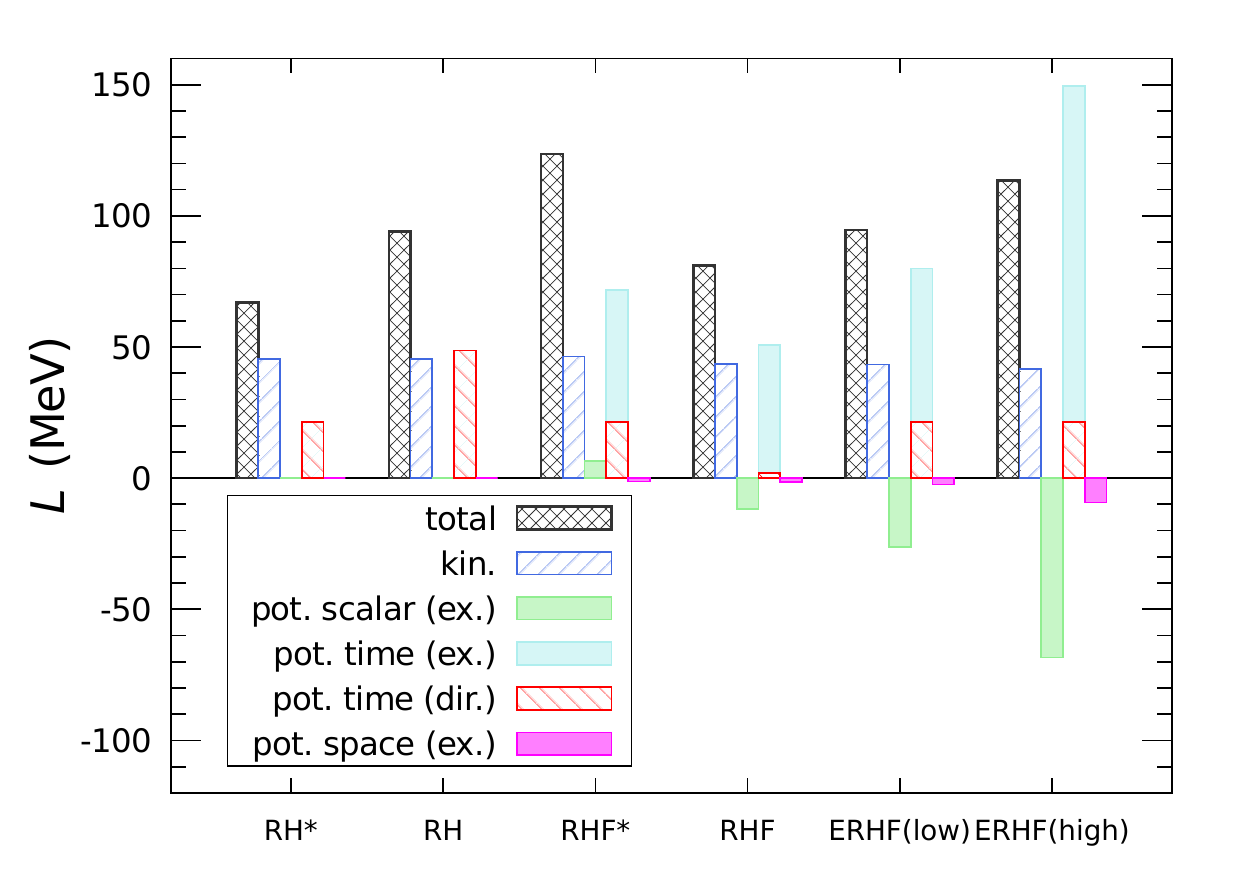}
\caption{\label{fig:L-rho0}
Components of $L$ at $\rho_{0}$. See the text for details.
}
\end{figure}
%%%%%%%%%%%%%%%%%%%%%%%%%%%%%%%%%%%%%%%%%%%%%%%%%%%%%%%%%%%%%%%%%%%%%%%%%%%%%%%%
The slope parameter of nuclear symmetry energy, $L$, at $\rho_{0}$ is shown in Fig.~\ref{fig:L-rho0}.
As in the case of $E_{\rm sym}$, $L$ is decomposed into the kinetic and potential terms, $L^{\rm kin}$ and $L^{\rm pot}$, and $L^{\rm pot}$ is again constructed by the three Lorentz components.
We find that $L^{\rm kin}$ shows 42--46 MeV in all the cases.
In $L^{\rm pot}$, although the direct term plays an important role in the RH and RH$^{\ast}$ models, the exchange contributions due to the scalar and time components dominate in the RHF and RHF$^{\ast}$ models.
In the ERHF models, $L^{\rm pot}$ itself is increased by the enhanced Fock contribution  even when $E_{\rm sym}$ is kept to be the appropriate value at $\rho_{0}$, and, as a consequence, $L$ is slightly far from the current global average based on the analyses of terrestrial nuclear experiments and astrophysical observations~\cite{Li:2013ola}.

%%%%%%%%%%%%%%%%%%%%%%%%%%%%%%%%%%%%%%%%%%%%%%%%%%%%%%%%%%%%%%%%%%%%%%%%%%%%%%%%
\begin{figure}
\includegraphics[width=15.0cm,keepaspectratio,clip]{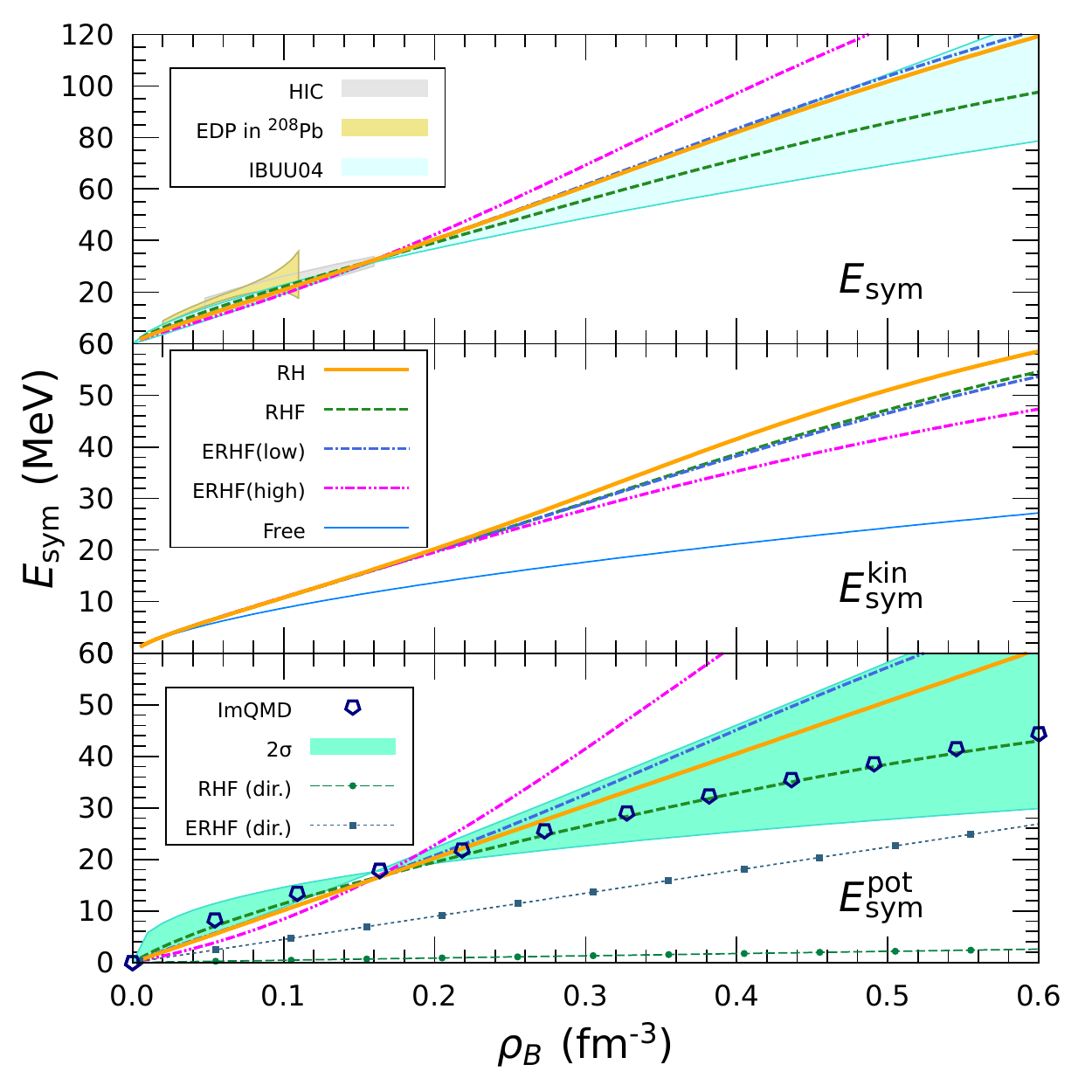}%
\caption{\label{fig:Esym-density}
Nuclear symmetry energy as a function of $\rho_{B}$.
The top panel is for the total nuclear symmetry energy, $E_{\rm sym}$.  The constraints from heavy-ion collisions (HIC) and experiments of electric dipole polarizability (EDP) in $^{208}$Pb are presented as well~\cite{Tsang:2012se,Zhang:2015ava}.
The result calculated by the isospin-dependent Boltzmann-Uehling-Uhlenbeck (IBUU04) transport model is also depicted using $E_{\rm sym}=31.6\left(\rho_{B}/\rho_{0}\right)^{x}$ with $x=$0.69--1.05~\cite{Chen:2004si,Li:2005jy}.
The middle (bottom) panel is for the kinetic (potential) term of $E_{\rm sym}$.
In the middle panel, ``free'' denotes $E_{\rm sym}^{\rm kin}$ in the case where the interactions are switched off.
The result based on the improved quantum molecular dynamics (ImQMD) transport model is also shown including $2\sigma$ confidence region in the bottom panel~\cite{Tsang:2008fd}.
}
\end{figure}
%%%%%%%%%%%%%%%%%%%%%%%%%%%%%%%%%%%%%%%%%%%%%%%%%%%%%%%%%%%%%%%%%%%%%%%%%%%%%%%%
The density dependence of nuclear symmetry energy, $E_{\rm sym}$, together with its kinetic and potential terms is presented in Fig.~\ref{fig:Esym-density} (see Eq.~\eqref{eq:Esym}).
The present results of $E_{\rm sym}$ are consistent with the constraints from heavy-ion collisions (HIC) and experiments of electric dipole polarizability (EDP) in $^{208}$Pb~\cite{Tsang:2012se,Zhang:2015ava}.
At the same time, in the RH, RHF, and ERHF(low) models, our results lie in the region of the isospin-dependent Boltzmann-Uehling-Uhlenbeck (IBUU04) transport calculation~\cite{Chen:2004si,Li:2005jy}.
In the ERHF(high) model, $E_{\rm sym}$ is larger than the IBUU04 result at densities above $\rho_{0}$.

The kinetic term of nuclear symmetry energy, $E_{\rm sym}^{\rm kin}$, is shown in the middle panel of Fig.~\ref{fig:Esym-density}.
We also show a line for the {\it free} kinetic term, in which the interactions are ignored, i.e.~$E_{\rm sym}^{\rm kin, free}=\frac{1}{6}\frac{k_{F}^{2}}{\sqrt{k_{F}^{2}+M_{N}^{2}}}$.
Owing to the relativistic many-body interactions, $E_{\rm sym}^{\rm kin}$ is larger than $E_{\rm sym}^{\rm kin, free}$ even at low densities (see Eq.~\eqref{eq:Esym-kin}).
Because the in-medium nucleon mass decreases most rapidly in the RH model, $E_{\rm sym}^{\rm kin}$ in RH approximation is larger than those in the other models.
In the RHF and ERHF models, the Fock terms suppress $E_{\rm sym}^{\rm kin}$ at densities above $\rho_{0}$.
As seen in the figure, at high densities, the larger the exchange contribution is, the smaller $E_{\rm sym}^{\rm kin}$ is.

As already explained in Fig.~\ref{fig:Esym-rho0}, the potential term of nuclear symmetry energy, $E_{\rm sym}^{\rm pot}$, can be divided into the direct and exchange parts: $E_{\rm sym}^{\rm pot}=E_{\rm sym}^{\rm pot,dir}+E_{\rm sym}^{\rm pot,ex}$.
The direct part, $E_{\rm sym}^{\rm pot,dir}$, is derived from the time component of the nucleon self-energy, and depends only upon the $\rho$-$N$ coupling constant.
In the bottom panel of Fig.~\ref{fig:Esym-density}, we present $E_{\rm sym}^{\rm pot,dir}$ as well as $E_{\rm sym}^{\rm pot}$.
In the RH model, $E_{\rm sym}^{\rm pot}$ is proportional to $\rho_{B}$, while $E_{\rm sym}^{\rm pot}$ does not increase linearly if the Fock contribution is taken into account.
In the RHF model, $E_{\rm sym}^{\rm pot}$ is mainly generated by $E_{\rm sym}^{\rm pot,ex}$, because $g_{\rho}$ is small and hence $E_{\rm sym}^{\rm pot,dir}$ too.
The present result of $E_{\rm sym}^{\rm pot}$ in RHF approximation is consistent with the constraint from analysis of heavy-ion collision data using the improved Quantum Molecular Dynamics (ImQMD) transport model~\cite{Tsang:2008fd}.
The energy difference between the low and high cases in the ERHF models is caused by the magnitude of $E_{\rm sym}^{\rm pot,ex}$, since $E_{\rm sym}^{\rm pot,dir}$ is identical in both cases.
Thus, it implies that the large exchange contribution enhances $E_{\rm sym}^{\rm pot}$ at high densities.

%%%%%%%%%%%%%%%%%%%%%%%%%%%%%%%%%%%%%%%%%%%%%%%%%%%%%%%%%%%%%%%%%%%%%%%%%%%%%%%%
\begin{figure}
\includegraphics[width=11.0cm,keepaspectratio,clip]{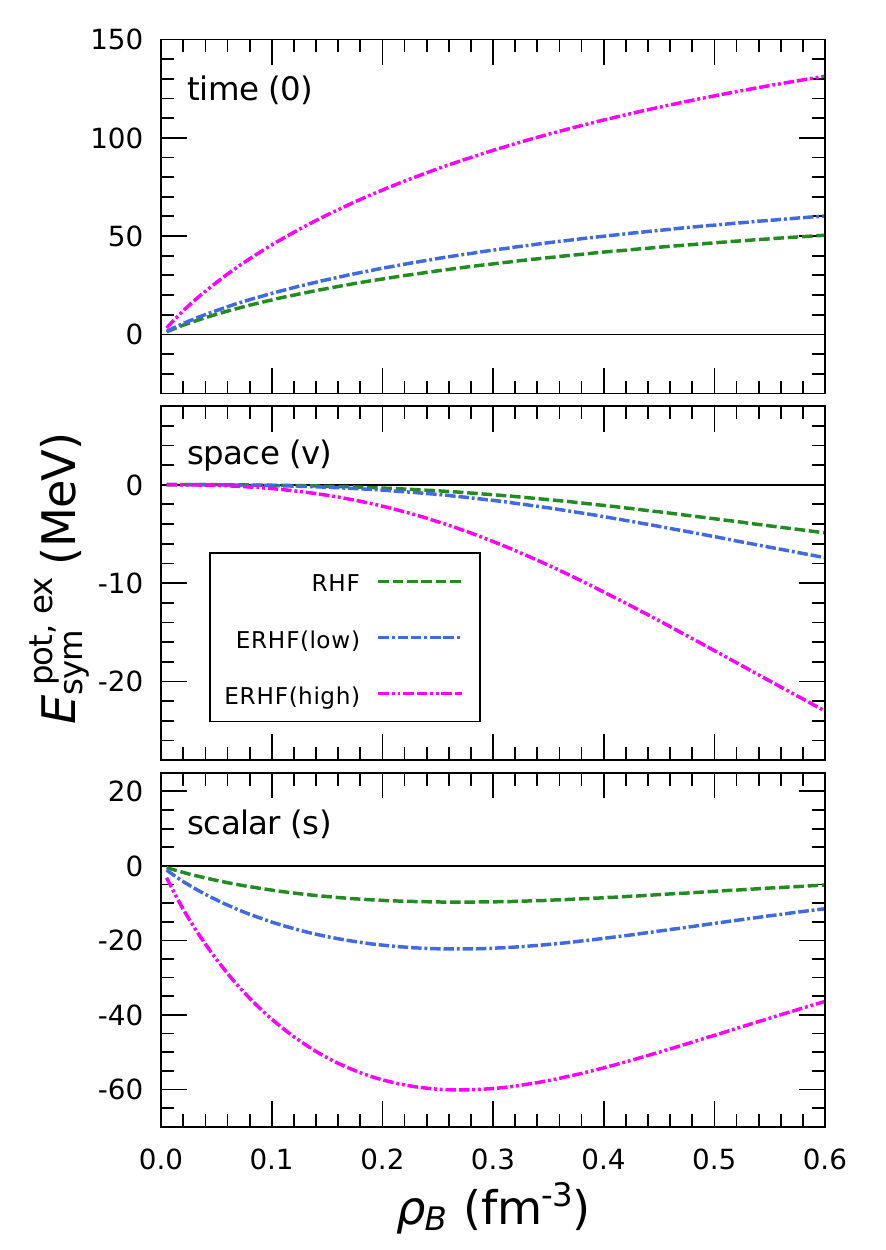}%
\caption{\label{fig:Esym-pot-ex}
Lorentz-covariant decomposition of $E_{\rm sym}^{\rm pot,ex}$ as a function of $\rho_{B}$.
The top (middle) [bottom] is for the time (space) [scalar] component of $E_{\rm sym}^{\rm pot,ex}$.
}
\end{figure}
%%%%%%%%%%%%%%%%%%%%%%%%%%%%%%%%%%%%%%%%%%%%%%%%%%%%%%%%%%%%%%%%%%%%%%%%%%%%%%%%
In Fig.~\ref{fig:Esym-pot-ex}, we present the components of $E_{\rm sym}^{\rm pot,ex}$ based on the Lorentz-covariant decomposition of $\Sigma_{N}$.
As in Fig.~\ref{fig:NSE-momentum}, $E_{\rm sym}^{\rm pot,ex}$ around $\rho_{0}$ is mainly determined by a sum of the scalar and time components.
With growing the density, the contribution due to the time and space components becomes large, while the effect of the scalar component turns to be relatively small.
As seen in Eq.~\eqref{eq:Esym-pot}, this is because the effective nucleon mass becomes considerably small and the effective momentum of nucleon increases at high densities.
It is also interesting to note that, although the space component is quite small at low densities and is often ignored in the DBHF calculation~\cite{Brockmann:1990cn}, it may be no longer negligible when we consider $E_{\rm sym}$ at high densities~\cite{Katayama:2013zya,Katayama:2015dga}.

%%%%%%%%%%%%%%%%%%%%%%%%%%%%%%%%%%%%%%%%%%%%%%%%%%%%%%%%%%%%%%%%%%%%%%%%%%%%%%%%
\begin{figure}% [h!]
\includegraphics[width=11.0cm,keepaspectratio,clip]{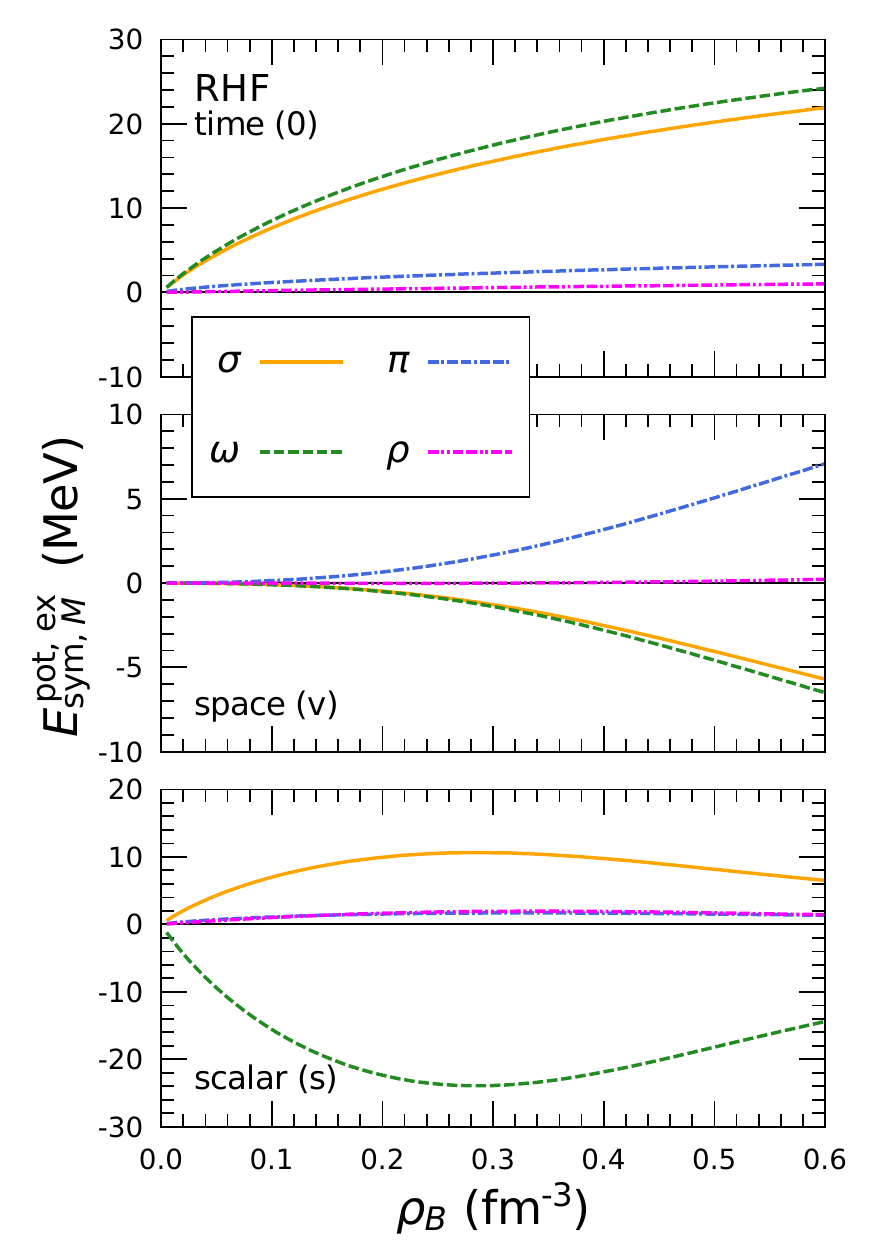}%
\caption{\label{fig:Esym-pot-ex-meson}
Density dependence of the meson contributions to the Lorentz-covariant components in $E_{\rm sym}^{\rm pot,ex}$.
This calculation is performed in RHF approximation.
}
\end{figure}
%%%%%%%%%%%%%%%%%%%%%%%%%%%%%%%%%%%%%%%%%%%%%%%%%%%%%%%%%%%%%%%%%%%%%%%%%%%%%%%%
In addition, in Fig.~\ref{fig:Esym-pot-ex-meson}, we show how the mesons ($M=\sigma,\omega,\rho,\pi$) contribute to $E_{\rm sym}^{\rm pot,ex}$ in the RHF model.
In RH approximation, $E_{\rm sym}^{\rm pot}$ ($=E_{\rm sym}^{\rm pot,dir}$) is affected only by the $\rho$ meson.
In contrast, in the RHF model, not only $\rho$ meson but also $\sigma$, $\omega$, and $\pi$ mesons influence $E_{\rm sym}^{\rm pot,ex}$.
It is thus of great interest that the $\sigma$ and $\omega$ mesons play an important role in $E_{\rm sym}^{\rm pot,ex}$.
On the other hand, the contribution due to the $\rho$ and $\pi$ mesons is extremely small even at high densities.
We note that this tendency can be seen in the ERHF models as well.

%%%%%%%%%%%%%%%%%%%%%%%%%%%%%%%%%%%%%%%%%%%%%%%%%%%%%%%%%%%%%%%%%%%%%%%%%%%%%%%%
%% Table: Esym-approximate
%%%%%%%%%%%%%%%%%%%%%%%%%%%%%%%%%%%%%%%%%%%%%%%%%%%%%%%%%%%%%%%%%%%%%%%%%%%%%%%%
\begin{table}%[h!]
\caption{\label{tab:Esym-approximate}
Fitting of $E_{\rm sym}$ using the free Fermi-gas-model formula.
The kinetic and potential values at $\rho_{0}$, $E_{\rm sym}^{\rm kin}(\rho_{0})$ and $E_{\rm sym}^{\rm pot}(\rho_{0})$, are calculated by each model, and they are in MeV.
For the sake of comparison, we show the constraints from (1) heavy-ion collision data using the ImQMD transport model~\cite{Tsang:2008fd}, (2) FOPI/LAND and ASY-EOS experiments using the the ultrarelativistic QMD model~\cite{Russotto:2011hq,Russotto:2016ucm}, and (3) astrophysical observations of neutron stars (NS-EOS)~\cite{Steiner:2010fz}.
The result of the correlated Fermi gas (CFG) model is also presented~\cite{Hen:2014yfa}.
}
\begin{ruledtabular}
\begin{tabular}{lccc}
\                                               & $E_{\rm sym}^{\rm kin}(\rho_{0})$ & $E_{\rm sym}^{\rm pot}(\rho_{0})$ & $\gamma$  \\
\hline
RH                                              &                              16.3 &                              16.2 &      1.00 \\
RHF                                             &                              16.1 &                              16.4 &      0.74 \\
ERHF(low)                                       &                              16.1 &                              16.4 &      1.09 \\
ERHF(high)                                      &                              16.1 &                              16.4 &      1.45 \\
\hline
(1) ImQMD~\cite{Tsang:2008fd}                   &                              12.5 &                              17.6 &      0.7$^{+0.35}_{-0.3}$ \\
(2) FOPI/LAND~\cite{Russotto:2011hq}            &                              12   &                              22   &      0.9$\pm0.4$          \\
\hspace*{0.5cm} ASY-EOS~\cite{Russotto:2016ucm} &                              12   &                              22   &      0.72$\pm0.19$        \\
(3) NS-EOS~\cite{Steiner:2010fz}                &                              17   &                            12--18 &      0.26 \\
\hline
CFG~\cite{Hen:2014yfa}                          &                        $-10\pm3$  &                              41   &      0.25 \\
\end{tabular}
\end{ruledtabular}
\end{table}
%%%%%%%%%%%%%%%%%%%%%%%%%%%%%%%%%%%%%%%%%%%%%%%%%%%%%%%%%%%%%%%%%%%%%%%%%%%%%%%%
In order to compare the present calculation with the phenomenological analyses on $E_{\rm sym}$, we here introduce the free Fermi-gas-model formula to describe the density dependence of $E_{\rm sym}$:
\begin{equation}
  E_{\rm sym}(\rho_{B}) \simeq E_{\rm sym}^{\rm kin}(\rho_{0}) \left(\rho_{B}/\rho_{0}\right)^{2/3}
                        +      E_{\rm sym}^{\rm pot}(\rho_{0}) \left(\rho_{B}/\rho_{0}\right)^{\gamma},
  \label{eq:Esym-approximate}
\end{equation}
which has been often used in the analyses in Refs~\cite{Zhang:2007qv,Sammarruca:2017edz}.
Using this function, in Table~\ref{tab:Esym-approximate}, we provide the $\gamma$ parameter in $E_{\rm sym}$, which is chosen so as to reproduce the result of each model.  For reference, several experimental analyses are also shown in the table.
%The fitting for the present models is performed by searching $\gamma$ with $E_{\rm sym}^{\rm kin}(\rho_{0})$ and $E_{\rm sym}^{\rm pot}(\rho_{0})$ which are the calculated values at $\rho_{0}$.
%Owing to the exchange contribution, $E_{\rm sym}$ can be reproduced with $\gamma=0.74$--1.45, in comparison with the result in the RH model, where $\gamma=1.00$.

We can see that the values of $\gamma$ in the present models lie within 0.74 -- 1.45, and they are consistent with those from heavy-ion collision data in the ImQMD transport model~\cite{Tsang:2008fd} and the ultrarelativistic QMD model~\cite{Russotto:2011hq,Russotto:2016ucm}.
However, the present values are much larger than the results obtained from the neutron-star analysis (NS-EOS)~\cite{Steiner:2010fz} and the Fermi gas model with the short-range correlations induced by the tensor force (CFG)~\cite{Hen:2014yfa}.
According to the recent calculations in RH approximation~\cite{Dutra:2017ysk}, the $\gamma$ parameter becomes smaller than unity if the nonlinear terms, $\bar{\sigma}\bar{\rho}^{2}$, $\bar{\omega}^{2}\bar{\rho}^{2}$, etc., as well as $\bar{\sigma}^{3}$ and $\bar{\sigma}^{4}$ are considered.
It may imply that higher-order terms of meson self-interactions at the Hartree level can imitate the effect of Fock terms.

% We give caveat.
% Although the approximate form given in Eq.~\eqref{eq:Esym-approximate} is available for the $E_{\rm sym}^{\rm pot}$ in the density region below maximally around $3\rho_{0}$, the $E_{\rm sym}^{\rm kin}$ can not be described adequately even at $\rho_{0}$.
% Thus, this approximate form may be no longer the effective parametrization theoretically, considering the density-dependent effect of the $E_{\rm sym}$ at supra-high densities~\cite{Sammarruca:2017edz,Brown:2013mga}.

%%%%%%%%%%%%%%%%%%%%%%%%%%%%%%%%%%%%%%%%%%%%%%%%%%%%%%%%%%%%%%%%%%%%%%%%%%%%%%%%
\begin{figure}% [h!]
\includegraphics[width=15.0cm,keepaspectratio,clip]{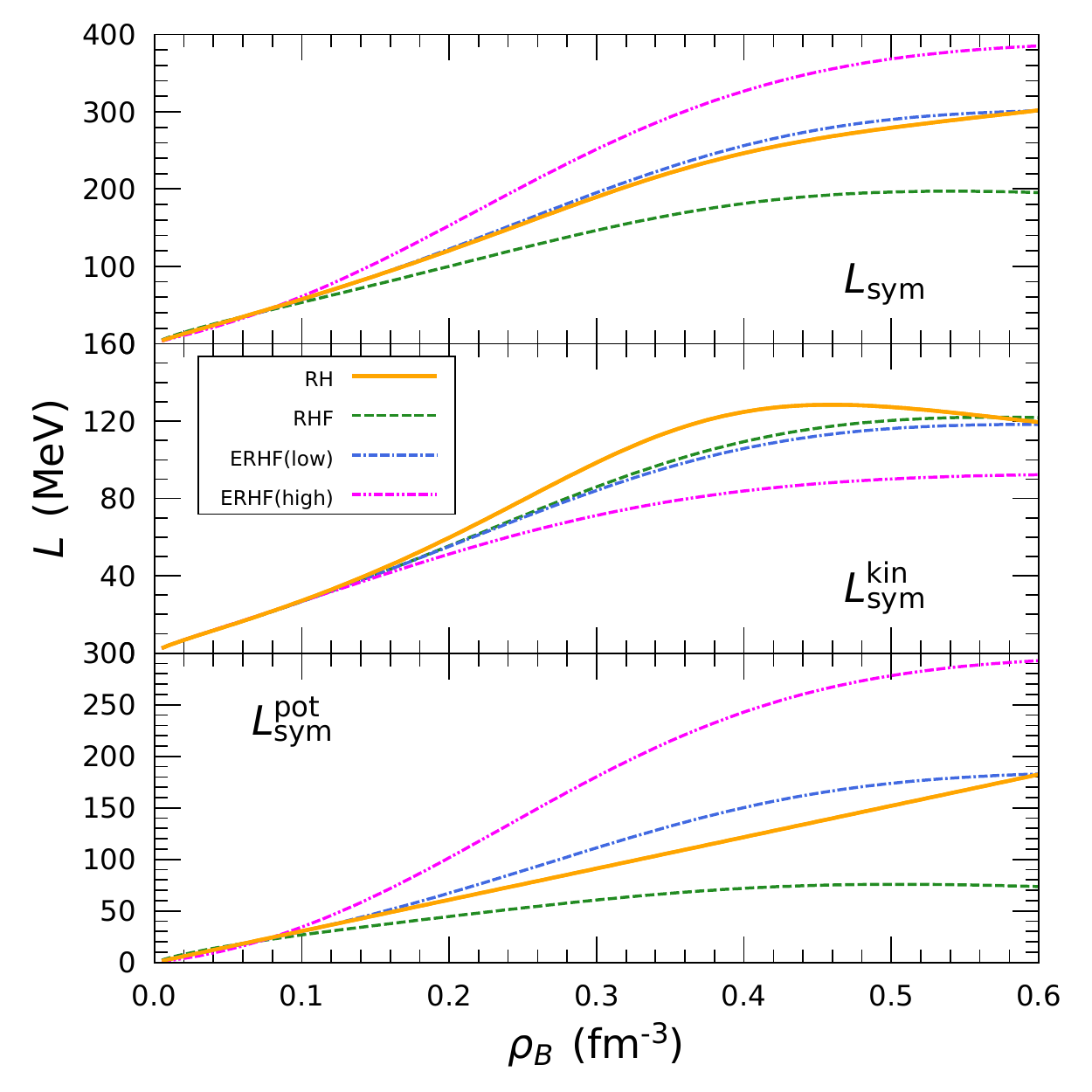}%
\caption{\label{fig:L-density}
Slope parameter of nuclear symmetry energy as a function of $\rho_{B}$.
The top panel is for the slope parameter, and the middle (bottom) one is for the kinetic (potential) term, $L^{\rm kin} (L^{\rm pot})$.
}
\end{figure}
%%%%%%%%%%%%%%%%%%%%%%%%%%%%%%%%%%%%%%%%%%%%%%%%%%%%%%%%%%%%%%%%%%%%%%%%%%%%%%%%
The density dependence of slope parameter, $L$, together with its kinetic and potential terms is presented in Fig.~\ref{fig:L-density}.
In the RH model, $L^{\rm kin}$ reaches a local maximum value around 0.45 fm$^{-3}$, then decreases for a while, and gradually grows like $k_{F}/6$ in the limit of $\rho_{B}\to\infty$ ($M_{N}^{\ast}\to0$) (see Eq.~\eqref{eq:slope-kin-dir}).
In contrast, $L^{\rm pot}$ is proportional to $\rho_{B}$ at high densities.
As a consequence, $L$ keeps on increasing monotonically in the RH model.

On the other hand, in the RHF and ERHF models, since both $L^{\rm kin}$ and $L^{\rm pot}$ increase and reach plateaus, respectively, with increasing $\rho_{B}$, $L$ is saturated at $3$ -- $4\rho_{0}$.
Because, as shown in Fig.~\ref{fig:Esym-density}, $E_{\rm sym}^{\rm pot, dir}$ is considerably small in the RHF model, the Hartree term due to the $\rho$ meson has little influence on $L^{\rm pot}$ even at high densities.
Therefore, at very high densities, $L$ in the RHF model becomes smaller than that in the RH model.
We note that, because the Fock contribution is enhanced as shown in~\ref{subsec:self-engy-USEP}, $L$ in the ERHF(high) model is larger than those in the other models below $\rho_{B}=0.6$ fm$^{-3}$.

% Maruyama san's comment:
% Finally, we leave a comment here.
% Apparently, Figs. 11 and 12 do (is) not consistent to the density-dependence given by Eq.(45).
% Then, we cannot estimate the symmetry energy in high density region with the values of L and Ksym.
% I thought , we agreed that we gave comments on that.

%%%%%%%%%%%%%%%%%%%%%%%%%%%%%%%%%%%%%%%%%%%%%%%%%%%%%%%%%%%%%%%%%%%%%%%%%%%%%%%%%%%%%%%%%%%%%%%%%%%%
\section{Summary}
\label{sec:summary}
%%%%%%%%%%%%%%%%%%%%%%%%%%%%%%%%%%%%%%%%%%%%%%%%%%%%%%%%%%%%%%%%%%%%%%%%%%%%%%%%%%%%%%%%%%%%%%%%%%%%

We have studied the effect of Fock terms on the properties of nuclear matter in a relativistic framework.
In particular, by taking into account the Lorentz-covariant decomposition of nucleon self-energies, we have discussed how the exchange contribution affects the nuclear symmetry energy, $E_{\rm sym}$,  and its slope parameter, $L$, up to $4\rho_{0}$.
In the present calculation, the relativistic Hartree (RH) and Hartree-Fock (RHF) models based on QHD have been employed to reveal the role of Fock contributions.
Furthermore, we have extended the RHF model to a new one, in which the Fock diagram is treated independently of the Hartree one, to investigate the relationship between the momentum dependence of nucleon self-energies and the properties of asymmetric nuclear matter.  We call it the ERHF model.

Using the Hugenholtz--Van Hove theorem, we have shown the analytical formula of $E_{\rm sym}$, which are divided into the kinetic and potential terms, $E_{\rm sym}^{\rm kin}$ and $E_{\rm sym}^{\rm pot}$.
In addition, we have studied $E_{\rm sym}^{\rm pot}$ in detail by separating it into the direct and exchange contributions, $E_{\rm sym}^{\rm pot,ex}$ and $E_{\rm sym}^{\rm pot,dir}$.
Then, as in the case of nucleon self-energies, $E_{\rm sym}^{\rm pot,ex}$ consists of the scalar, time and space components based on the Lorentz-covariant decomposition.
Moreover, we have calculated $L$, which is also separated into the kinetic and potential terms as $L=L^{\rm kin}+L^{\rm pot}$.

We here summarize our results as follows:
\begin{enumerate}
\item We have demonstrated the momentum dependence of nucleon self-energies.
The scalar and time components mainly compose nucleon self-energies, while the space component is negligibly small around $\rho_{0}$.
Furthermore, we have studied the single-nucleon potential using the Dirac phenomenology.
Because of the momentum dependence through the nucleon self-energies, in the ERHF model, it is possible to reproduce the experimental scattering data without introducing any density-dependent couplings which are generally required in the RMF models~\cite{Typel:2005ba} (see Figs.~\ref{fig:NSE-momentum} and \ref{fig:USEP}).
\item As for the properties of dense nuclear matter, we have calculated the effective nucleon mass, nuclear binding energy and pressure.
It has been found that Fock terms suppress the rapid reduction of effective nucleon mass at densities beyond $\rho_{0}$.
It is noticeable that the exchange contribution makes pressure soft at high densities in both cases of symmetric nuclear and pure neutron matter (see Figs.~\ref{fig:Emass}--\ref{fig:Pressure}).
\item In the RH, RHF and ERHF models, if $E_{\rm sym}$ is chosen to be the empirical value at $\rho_{0}$, namely $E_{\rm sym}=32.5$ MeV, $E_{\rm sym}^{\rm kin}$ and $E_{\rm sym}^{\rm pot}$ are about 16 and 16.5 MeV, respectively. Furthermore, the exchange contribution dominates in $E_{\rm sym}^{\rm pot}$, and $E_{\rm sym}^{\rm pot,ex}$ is estimated to be 15.7 (9.3) MeV in the RHF (ERHF) model.
Concerning $L$ at $\rho_{0}$, $L^{\rm kin}$ is 42 -- 46 MeV in all the models.
The exchange contribution suppresses $L^{\rm pot}$ in the RHF model, while, in the ERHF model, the enhanced Fock contribution makes $L^{\rm pot}$ larger than that in the RH model (see Figs~\ref{fig:Esym-rho0} and \ref{fig:L-rho0}).
\item Even in the density region above $\rho_{0}$, we have found that, in general, Fock terms suppress $E_{\rm sym}^{\rm kin}$.
In the RHF model, $E_{\rm sym}^{\rm pot}$ is pushed downward by the Fock contribution and, thus, $E_{\rm sym}$ is smaller than that in the RH model.
On the other hand, in the ERHF model, $E_{\rm sym}^{\rm pot}$ and $E_{\rm sym}$ become larger than those in the RH model, which is caused by the enhanced exchange contribution.
Although the properties of asymmetric nuclear matter are often described by introducing only the $\rho$ meson in the RH models, the $\sigma$, $\omega$, and $\pi$ mesons as well as $\rho$ meson have considerable influence on $E_{\rm sym}$ through Fock terms.
In particular, in RHF approximation, the $\rho$ meson contribution to $E_{\rm sym}^{\rm pot,dir}$ is smaller than that in RH approximation, and the exchange terms due to the $\sigma$ and $\omega$ mesons contribute to $E_{\rm sym}^{\rm pot,ex}$ considerably.
As a consequence, $E_{\rm sym}^{\rm pot}$ in the RHF model becomes consistent with the constraint from heavy-ion collision data with the ImQMD transport model~\cite{Tsang:2008fd}.
Furthermore, since, with increasing the density, both $L^{\rm kin}$ and $L^{\rm pot}$ are suppressed by the exchange contribution, $L$ reaches a constant value at $3$ -- $4\rho_{0}$ in the RHF and ERHF models (see Figs~\ref{fig:Esym-density}, \ref{fig:Esym-pot-ex-meson}, and \ref{fig:L-density}).
\end{enumerate}

Finally, we comment on future works.
In the present study, we do not consider the short-range nucleon-nucleon correlations (SRC) in matter, which may contribute to $E_{\rm sym}$ and $L$.
It has been reported that the SRC due to the tensor force between a neutron-proton pair significantly increases the high momentum tail in symmetric nuclear matter, and $E_{\rm sym}^{\rm kin}$ then decreases even at $\rho_{0}$~\cite{Hen:2014yfa,Li:2014vua,Cai:2015xga}.
Thus, it may be important to consider the effect of high momentum component of nucleon induced by the SRC~\cite{Brockmann:1990cn,Katayama:2013zya,Katayama:2015dga}.

In the present calculation, nucleons are treated as point-like objects, and we do not pay any attention to the effect of quark degrees of freedom inside a nucleon.
In fact, the evidence for the medium modification of nucleon structure in a nucleus has been observed in polarization transfer measurements in the quasi-elastic $(e,e^{\prime}p)$ reaction at the Thomas Jefferson Laboratory~\cite{Brooks:2011sa}.
Furthermore, the recent lattice simulation has also suggested the nucleon structure change in medium~\cite{Chang:2017eiq}.
These results support the prediction of the quark-meson coupling (QMC) model~\cite{Guichon:1987jp,Saito:1994ki,Saito:2005rv,Nagai:2008ai}.
We also note that, in Refs.~\cite{Haiden:2019} and \cite{Machleidt:2019}, the importance of non-local interactions in nuclear force has been discussed recently.
Therefore, it is also interesting to study how quarks in a nucleon affects the matter properties including $E_{\rm sym}$ and $L$~\cite{Panda:2011sp,Providencia:2013cea}.

%%%%%%%%%%%%%%%%%%%%%%%%%%%%%%%%%%%%%%%%%%%%%%%%%%%%%%%%%%%%%%%%%%%%%%%%%%%%%%%%%%%%%%%%%%%%%%%%%%%%
\begin{acknowledgments}
This work was supported by JSPS KAKENHI Grant Numbers JP16K05360, JP17K14298.
The work of MKC was supported by the National Research Foundation of Korea (Grant No.~NRF-2017R1E1A1A01074023).
The work of KSK was supported by the National Research Foundation of Korea (MSIT No.~2018R1A5A1025563).
\end{acknowledgments}
%%%%%%%%%%%%%%%%%%%%%%%%%%%%%%%%%%%%%%%%%%%%%%%%%%%%%%%%%%%%%%%%%%%%%%%%%%%%%%%%%%%%%%%%%%%%%%%%%%%%

%%%%%%%%%%%%%%%%%%%%%%%%%%%%%%%%%%%%%%%%%%%%%%%%%%%%%%%%%%%%%%%%%%%%%%%%%%%%%%%%%%%%%%%%%%%%%%%%%%%%
% \appendix
% \section{Appendix}
% \label{sec:appendix}
%%%%%%%%%%%%%%%%%%%%%%%%%%%%%%%%%%%%%%%%%%%%%%%%%%%%%%%%%%%%%%%%%%%%%%%%%%%%%%%%%%%%%%%%%%%%%%%%%%%%

%

%
\end{document}